\documentclass[useAMS,usenatbib,pdfsync]{mn2e}
\usepackage{graphicx}

\newcommand       \be           {\begin{equation}}
\newcommand       \ee           {\end{equation}}
\newcommand       \bea          {\begin{eqnarray}}
\newcommand       \eea          {\end{eqnarray}}
\newcommand       \apj          {ApJ}
\newcommand       \apjl         {ApJL}
\newcommand       \aap          {A\&A}
\newcommand       \nat          {Nature}
\newcommand       \mnras        {MNRAS}
\def\simlt{\mathrel{\hbox{\rlap{\hbox{\lower4pt\hbox{$\sim$}}}\hbox{$<$}}}}
\def\simgt{\mathrel{\hbox{\rlap{\hbox{\lower4pt\hbox{$\sim$}}}\hbox{$>$}}}}

\title[Magnetar Jets and GRBs]{Magnetized Relativistic Jets and
  Long-Duration GRBs from Magnetar Spindown during Core-Collapse Supernovae}

\author[N. Bucciantini, at al.]{N. Bucciantini$^{1}$\thanks{E-mail:
    nbucciantini@astro.berkeley.edu}, E. Quataert$^{1,2}$, B.~D. Metzger$^{1,2}$,
  T.~A. Thompson$^{3}$,\newauthor J. Arons$^{1,2}$, L. Del Zanna$^{4}$\\
  $^{1}$Astronomy Department and Theoretical Astrophysics Center,
  University of California, Berkeley, 601 Campbell Hall,\\ Berkeley CA,
  94720\\ $^{2}$Department of Physics, University of California,
  Berkeley, Le Conte Hall, Berkeley, CA 94720\\
  $^{3}$Department of Astronomy and Center for Cosmology \& Astro-Particle
 Physics, The Ohio State University, Columbus,
 Ohio, 43210\\ $^{4}$Dipartimento di Astronomia e Scienza dello Spazio,
  Universit\`a di Firenze, L.go Fermi 2, 50125, Firenze, Italy}
\begin{document}

\date{Accepted . Received ; in original form }

\pagerange{\pageref{firstpage}--\pageref{lastpage}} \pubyear{????}

\maketitle

\label{firstpage}

\begin{abstract}

  We use ideal axisymmetric relativistic magnetohydrodynamic
  simulations to calculate the spindown of a newly formed millisecond,
  $B \sim 10^{15}$ G, magnetar
  and its interaction with the surrounding stellar envelope during a
  core-collapse supernova (SN) explosion.  The mass, angular momentum,
  and rotational energy lost by the neutron star are determined
  self-consistently given the thermal properties of the cooling
  neutron star's atmosphere and the wind's interaction with the
  surrounding star.  The magnetar drives a relativistic magnetized
  wind into a cavity created by the outgoing SN shock. For high
  spindown powers ($\sim 10^{51}-10^{52}$ ergs s$^{-1}$), the magnetar
  wind is super-fast at almost all latitudes, while for lower spindown
  powers ($\sim 10^{50}$ ergs s$^{-1}$), the wind is sub-fast but
  still super-Alfv\'enic.  In all cases, the rates at which the
  neutron star loses mass, angular momentum, and energy are very
  similar to the corresponding free wind values ($\simlt 30 \%$
  differences), in spite of the causal contact between the neutron
  star and the stellar envelope.  In addition, in all cases that we
  consider, the magnetar drives a collimated ($\sim 5-10^\circ$)
  relativistic jet out along the rotation axis of the star.  Nearly
  all of the spindown power of the neutron star escapes via this polar
  jet, rather than being transferred to the more spherical SN
  explosion.  The properties of this relativistic jet and its expected
  late-time evolution in the magnetar model are broadly consistent
  with observations of long duration gamma-ray bursts (GRBs) and their
  associated broad-lined Type Ic SN.

\end{abstract}

\begin{keywords}
Stars: neutron; stars: supernovae: general; gamma-rays: bursts; stars: winds,
outflows; magnetic field; MHD
\end{keywords}

\section{Introduction}
\label{sec:int}

Observations of long-duration gamma-ray bursts (LGRBs) have
demonstrated that they are associated with core-collapse supernovae
(SNe) and the death of massive stars
(\citealt{woo06,del06,zha07}). Two leading candidates for the central
engine powering LGRBs are a newly formed, rapidly rotating magnetar
(e.g., \citealt{usov92,thom94,wheeler00,thom04}) or a black hole with
an accretion disk (e.g., \citealt{mcf99}). In both of these cases, it
is crucial to understand how relativistic material generated by the
central engine -- which is believed to give rise to the GRB at large
distances from the compact object -- escapes from deep within the host
star.  The most likely possibility, suggested by both afterglow
observations (``jet breaks''; \citealt{rho99}) and GRB energetics, is
that a collimated outflow punches through the stellar envelope,
providing a channel out of which the relativistic material can escape.

In the collapsar model, collimated outflows from GRBs are accounted
for by jets produced by a centrifugally supported accretion flow onto
a central black hole, based on the results of numerical simulations
(e.g., \citealt{proga03}) and by analogy with X-ray binary and AGN
jets.

In the magnetar model, however, the origin of such collimated outflows
is less clear because relativistic outflows  do not
efficiently self-collimate (e.g., \citealt{le01}). Numerical
simulations of outflows from {\it isolated} magnetars have clearly
shown that
a highly collimated outflow can only be formed at very early times
(within a few seconds after core bounce) when the outflow is only
mildly relativistic, while relativistic outflows can only be produced
at later times when the latitudinal distribution of the energy-flux
approaches that of the force-free solution, with most of the energy
flux confined in the equatorial region and not along the axis
\citep{me06}.  These results appear to be independent of the specific
magnetic field configuration on the surface of the star.

Energy considerations demonstrate, however, that the surrounding
stellar envelope provides an efficient confining medium even for a
very energetic proto-magnetar wind.  Thus, the GRB
outflow might be strongly affected by the interaction with the
progenitor star (this is true in both the collapsar and magnetar
contexts).  \citet{kg02} suggested, by analogy to pulsar wind nebulae
\citep{beg92} (PWNe), that the interaction of the wind from the
spinning-down magnetar with the surrounding star could facilitate
collimation.

In two previous papers (\citealt{b08a}, Paper I; \citealt{b08b}, Paper
II), we have investigated the dynamics of this interaction, using a
variety of simplifying assumptions. In Paper I we used axisymmetric
thin shell calculations to demonstrate that collimation can indeed
occur (see \citealt{um06,um07} for related ideas based on force-free
rather than inertially loaded outflows).  In Paper II we used
relativistic magnetohydrodynamic (MHD) simulations to investigate the
interaction of a relativistic magnetized wind with the progenitor
star, using winds with properties expected for proto-magnetars (based
on \citealt{met07}). By analogy with PWNe (\citealt{kom04,ldz04}), we
focused on the low-magnetization regime in which most (but not all) of
the magnetic energy of the wind is assumed to be converted into bulk
kinetic energy.  We found that the magnetar wind shocks on the
surrounding (exploding) stellar envelope, creating a bubble of
relativistic plasma and magnetic field inside the star (a ``magnetar
wind nebula''; MWN).  Just as in PWNe, if the toroidal magnetic field
in the bubble is sufficiently strong, the bubble expands primarily in
the polar direction due to the ``tube of toothpaste'' effect - the
magnetic hoop stress in the equator creates a pressure gradient along
the axis which is not balanced by any restoring force, thus driving
the flow preferentially in the polar direction. The nebula itself is
ultimately confined by the inertia of the SN ejecta, to which little
energy is transferred, in contrast to the pressure confinement in
traditional magnetic tower models (\citealt{she07}) or confinement by
a pressurized cocoon inside the progenitor star (\citealt{um06}).

The results of Paper I and II, despite being obtained with different
approaches (thin-shell semi-analytic calculations {\it vs.} full
relativistic MHD simulations) and utilizing different assumptions,
demonstrate qualitatively similar overall dynamics.  In both cases,
however, in order to reduce the computational time of the
calculations, the properties of the outflow from the central source
were specified using a fixed prescription, based on previous results
for free-flowing winds \citep{me06,met07}, and using
super-fastmagnetosonic injection.  One concern with this
approach is that, if the termination shock between the wind and the
surrounding star moves within the slow, Alfv\'en, or fast critical
surfaces, the stellar envelope is causally connected to the central
star, which may modify how the star spins down and loses rotational
energy and angular momentum (invalidating the boundary conditions used
at the central source).  In order to properly address these issues it
is necessary to follow the dynamics from the proto-magnetar atmosphere
out to large radii.

\cite{kom07} carried out such a simulation and found that there is
indeed some causal contact between the stellar envelope and the
central star.  In spite of this difference relative to our assumptions
in Papers I and II, the collimation of the outflow by magnetic hoop
stresses was similar to what we have previously found.  \cite{kom07}'s
simulations were only for $\sim 200$ ms after core bounce, however,
and thus only apply to the early non-relativistic phase of magnetar
spindown, not to the late-time relativistic, and potentially
GRB-producing, phase.

In this paper, we build on our previous work by carrying out
time-dependent axisymmetric relativistic MHD simulations of the
development of magnetar outflows and their propagation through a
surrounding star. We follow the dynamics of the proto-magnetar wind
into the late-time relativistic phase (for $\sim 10$ sec), and from
the injection radius located at the proto-neutron star surface to a
distance of a few progenitor stellar radii. Given the computational
requirements of this problem, which extends over several orders of
magnitude in radius, and which involves timescales ranging from
$10^{-7}$ sec in the proto-neutron star atmosphere to $\sim 10 $ sec
for jet propagation through the host star, we have not carried out a
detailed parameter study, but have instead limited our investigation
to a few fiducial cases.

As in Paper II, we assume that an outgoing SN shock has already
created a central evacuated cavity and that the host star is
spherically symmetric. However, unlike in Paper II, we make no direct
assumptions about the properties of the magnetar outflow.  We only set
the physical conditions at the proto-neutron star surface;
specifically, the density, temperature and the radial component of the
magnetic field are imposed there.  A transonic flow self-consistently
develops, with a speed increasing with radius. The wind magnetization
$\sigma=\Omega^2 \Phi^2/\dot{M}$ (where $\Omega$ is the rotation rate,
$\Phi$ the open magnetic flux, and $\dot{M}$ the mass loss rate)
changes in time as the mass loss rate drops due to the
changing temperature and density in the neutron star atmosphere (which
change as a result of the decreasing neutrino flux during the neutron
star's Kelvin-Helmholz phase).  Given that the outflow is not injected
with fixed properties, but is instead allowed to self-consistently
develop, any feedback due to the interaction with the progenitor star
is properly modeled.  As a result (and unlike in Paper II), we do not
assume that magnetic energy is converted into kinetic energy at large
radii in the wind.  On the contrary, the outflows simulated here are
always reasonably highly magnetized, from the early-time ($\simlt$ a
few sec) non-relativistic wind in which the magnetic energy is
comparable to the kinetic energy at large radii, to the late-time
magnetically dominated relativistic outflow.  It is, however, worth
noting that our calculations may not capture various instabilities
(e.g., 3D ones) that could dissipate magnetic energy, converting it
into thermal energy and ultimately bulk kinetic energy; this limit of
low $\sigma$ is explored in Paper II.

The remainder of this paper is organized as follows.  In \S
\ref{sec:num} we describe our numerical methodology.  We then describe
the results of our calculations, including their implications for the
collimation of magnetar outflows in LGRBs, the spindown of the central
neutron star, and nucleosynthesis of shock-heated stellar material (\S
\ref{sec:res}).  Finally, in \S \ref{sec:dis}, we summarize our
results and their implications.

\section{Numerical setup}
\label{sec:num}

All of the simulations were performed using ECHO, a shock-capturing
central-scheme for general relativistic ideal MHD; see \citet{ldz02}
and \citet{ldz03,ldz07} for a detailed description of the equations
and numerical algorithms.

The interaction of the magnetar outflow with the surrounding SN
progenitor is investigated by performing 2D axisymmetric simulations
on a spherical grid. The domain in $\theta$ is the first quadrant from
$\theta = 0$ to $\theta = \pi/2$, with reflecting boundary conditions
for $v_\theta$, $v_\phi$ and the magnetic field components $B_\theta$
and $B_\phi$ at the polar axis to enforce axisymmetry, and similar
boundary conditions in the equatorial plane.  The grid is uniform in
the $\theta$ direction with 100 cells.  Given that we are studying a
wide range of spatial scales, from the proto-magnetar atmosphere at
$\sim 10^6$ cm to the outer edge of the star at $\sim 2\,10^{10}$ cm,
we have selected a logarithmic spacing in radius with an inner
boundary located at $r_{\min}=1.5 \, 10^6$ cm and an outer boundary at
$r_{\max}=6.5 \, 10^{10}$ cm, and 100 cells per decade in
radius. Zeroth order extrapolation is assumed at the outer boundary.
The code is second order in both space and time, with a monotonized
central limiter, chosen in order to resolve the large density jump
between the lighter relativistic plasma inside the MWN, associated
with the magnetar outflow, and the heavier stellar envelope (the
density can increase by a factor of $\sim 10^4$).  We use a
Schwarzschild metric to account for the gravity of the central
proto-neutron star, which must be included to properly drive a
transonic wind from the neutron star surface. We do not include the
self-gravity of the progenitor star; the typical dynamical timescale
for the progenitor is longer than the duration of our simulations and
so the gravity of the progenitor can be neglected to first
approximation.

%%%%%%%%%%%%%%%%%%%%%%%%%%%%%%%%% Table 1 %%%%%%%%%%%%%%%%%%%%%%%
\begin{table*}
\begin{minipage}{17cm}
  \caption{Comparison of the free-wind mass loss rates obtained using
    the isothermal layer approach described in \S \ref{sec:num}, with
    the values obtained by solving the full neutrino-heated MHD wind
    problem \citep{met07}, both using a 1D monopole magnetic
    field. The cases with Period = $\infty$ correspond to non-rotating neutron
    stars, and are indicative of polar outflows in our 2D
    simulations. Time is after core bounce and mass loss rate $\dot M$
    is in $M_\odot \, s^{-1}$.}
\label{table:1}
\begin{center}
\begin{tabular}{c c c c c}
\hline
 Time (s) & Period (ms) & Magnetic field (G) & $\dot M$ (\citealt{met07}) & $\dot M$ (this paper) \\
\hline
1. & $\infty$ & $10^{15}$ & $1.5\times 10^{-4}$ & $1.6\times 10^{-4}$  \\
1. & 1.     & $10^{15}$ & $1.2\times 10^{-2}$ & $1.6\times 10^{-2}$  \\
1. & 3.     & $10^{15}$ & $3.0\times 10^{-4}$ & $4.0\times 10^{-4}$  \\
2. & $\infty$ & $10^{15}$ & $3.0\times 10^{-5}$ & $3.1\times 10^{-5}$  \\
2. & 1.     & $10^{15}$ & $2.4\times 10^{-3}$ & $2.9\times 10^{-3} $ \\
8. & 1.     & $10^{15}$ & $9.3\times 10^{-5}$ & $9.0\times 10^{-5}$  \\
1. & 1.     & $10^{16}$ & $1.5\times 10^{-2}$ & $1.9\times 10^{-2}$  \\
2. & 1.     & $10^{16}$ & $2.6\times 10^{-3}$ & $3.0\times 10^{-3} $ \\
\hline
\end{tabular}
\end{center}
\end{minipage}
\end{table*}

%%%%%%%%%%%%%%%%%%%%%%%%%%%%%%%%%%%%%%%%%%%%%%%%%%%%%%%%%%%%%%%%%%%%%%%%

Our previous results (Papers I and II) imply that it takes $\sim 5-10$
s for the magnetar outflow to emerge from the progenitor surface and
accelerate into the circumstellar medium. For this reason, we follow
the evolution of the system for $\simeq 10$ s.  Note that most of the
times quoted in this paper are given in seconds {\it after core
  bounce}. Our simulations begin 1 second after core bounce, which is
approximatively the time it takes for the proto-neutron star to
contract to its final radius $\simeq 15$ km and for the neutrino
driven wind to develop inside the SN ejecta \citep{sch06}.  In order
to evolve the dynamics in the neutron star atmosphere at $1.5 \, 10^6$
cm, the CLF time-step must be $\simlt 3 \, 10^{-7}$s. This highlights
the fact that studying the evolution of the system for the desired
duration of $\sim 10$ s is very time consuming. As a consequence, we
have been forced to make several simplifications in our treatment of
the microphysics. In particular, we use an ideal gas equation of state
with an adiabatic index of 4/3, which is appropriate for both the
relativistic magnetar outflow and for the radiation pressure dominated
shocks that result due to the interaction with the surrounding star;
the validity of this assumption is checked in \S \ref{sec:nuc}.

The effects of neutrino heating and cooling, which are responsible for
driving the wind, are primarily confined to a very small region around
the proto-neutron star \citep{thompson01,met07}.  We approximate the
effect of neutrino heating by including an isothermal layer at small
radii near the inner boundary. At the inner radius we fix the rotation
rate $\Omega$ and the value of the radial magnetic field $B_r$, and we
assume perfect conductivity so that $E_\theta=E_\phi=0$. The
temperature, density, and radial extent of the isothermal layer have
been adjusted to reproduce reasonably accurately the mass loss rate
obtained in more sophisticated 1D calculations that include a detailed
treatment of the microphysics \citep{met07}, and to still guarantee
numerical stability (decreasing the temperature leads to numerical
instabilities, especially at high magnetization). We found that a
reasonable choice was for an isothermal layer extending from the
stellar surface to $3 \, 10^6$ cm with a pressure $p$ such that $\rho
c^2/p \simeq 100$ (i.e., a sound speed $c_s \simeq 0.1 \, c$).  The
density $\rho$ decreases in time $\propto t^{-2.7}$ in order to
reproduce the decline in mass loss rate for a free wind that is caused
by the decreasing neutrino luminosity of the neutron star in the first
10 sec after core bounce \citep{bur86}; the radial extent and the
sound speed (set by $p/\rho c^2$) in the isothermal layer are
independent of time.  Given that the wind is magnetocentrifugally
accelerated, the value of the pressure does not affect its asymptotic
dynamics (so long as the wind is cold). 

The evolution of the magnetization $\sigma$, and the mass, energy, and
angular momentum loss rates for a free wind are shown in
Fig.~\ref{fig:losses} (dashed line), discussed in \S \ref{sec:res}. A
comparison between the mass-loss rates obtained using the isothermal
layer approach described here with a 1D model including a full
treatment of the neutrino physics \citep{met07} is shown in
Table~\ref{table:1}, and in Fig.~\ref{fig:losses}.  The agreement is
very good.  Note also that the methodology adopted here produces the
correct latitudinal variation in the mass-loss rate (see
\citealt{me06}), with magnetocentrifugal support enhancing the
mass-loss rate at the equator by a factor of $\sim 100$ relative to
the pole for a neutron star with a ms rotation period.  Although we
have emphasized the comparison to free-wind calculations as a way of
calibrating our simplified model, it is important to stress that our
boundary conditions only specify the density, temperature, and
magnetic field in the neutron star's atmosphere; they do not specify
the mass, angular momentum, or energy loss rates, which are determined
self-consistently by the wind's dynamics.

As in Papers I \& II, we use the $35 M_\odot$ model from \citet{woo02}
as our progenitor star.  The outer surface of the progenitor is
located at $2.5\, 10^{10}$ cm. We assume that the density outside the
star falls off as $r^{-2}$ as expected for a wind.  We have previously
verified that for the range of radii we simulate, our results are
independent of the outer density profile (Paper II).  In order to
account for the effect of a SN shock propagating inside the
progenitor, the region between $10^9$ and $2\, 10^9$ cm is given an
initial outward velocity corresponding to a total kinetic energy $2\,
10^{51}$ erg, similar to that used in the one-dimensional explosion
calculations of \citet{ww95}. This corresponds to the SN shock moving
at $\simeq 1.5 \, 10^4$ km/s, 1 second after core bounce. Hydrodynamic
studies of neutrino driven winds after successful core-collapse SNe
show that as the SN shock expands inside the star a cavity is left at
small radii, inside which the wind blows \citep{thom05,sch06}. To
approximate this, our initial condition includes a cavity inside the
progenitor with a radius of $10^9$ cm, which is roughly the size of
the collapsing progenitor and the location of the SN shock $1$ sec
after core bounce.  The cavity is initially filled with low density
plasma in hydrostatic equilibrium. The density is chosen to be low
enough that the proto-magnetar wind can initially expand freely into
the cavity.  The initial conditions do not, however, determine whether
the wind can expand freely at later times; this depends on the
subsequent evolution, which is self-consistently calculated.

The boundary conditions used here remove many of the assumptions about
the wind structure that were used in our previous papers, and allow
the outflow to develop freely and self-consistently according to the
physics of the interaction between the wind and the progenitor
star. The magnetic field is assumed to be monopolar and to extend from
the surface of the proto-neutron star into the SN progenitor. This
formally corresponds to assuming a large fossil magnetic field in the
progenitor.  We expect, however, that our results would be very
similar if the field is generated in situ by a convective dynamo
\citep{duc92}.  For example, outside the light cylinder ($\simeq 10^7$
cm for a millisecond rotator), the structure and properties of the
wind depend only on the amount of open magnetic flux, and not on its
structure in the closed magnetosphere, so that the monopole solution
is reasonable.  Moreover, even if the poloidal magnetic field threads
the entire star, at $10^9$ cm (the initial size of the SN ejecta) it
is dynamically negligible, and is unable to exert any significant
torque on the central compact object.  For the typical values of
$\sigma$ in our simulations, the amount of open magnetic flux in the
wind implies that the dipolar magnetic field is a factor $\sim 2$
larger than the monopole values quoted here (\citealt{me06}).

To explore a few reasonable models for magnetar central engines of
LGRBs, we consider three cases, motivated by previous work (e.g.,
\citealt{met07}): Case A is a 1 millisecond rotator with a
$B_r=10^{15}G$ surface magnetic field; Case B is a 1 millisecond
rotator with $B_r= 3 \, 10^{15}G$; and Case C is a 3 millisecond
rotator with $B_r=10^{15}G$. In all cases the
rotation rate is kept constant in time.  After the fact, we can assess
that this is a reasonable assumption for both cases A and C, in that
only a modest fraction of the rotational energy is lost during the 10
seconds of our simulation (see Fig. \ref{fig:losses}, discussed
below). For case B, however, keeping the rotation rate constant is not
fully self-consistent, because of the higher energy loss rate. Case B
is nonetheless a useful guide to the dynamics in the high $\sigma$,
high spindown power limit.

%%%%%%%%%%%%%%%%%%%%%%%%%%%%%%%%%%%%%% FIG 1 %%%%%%%%%%%%%%%%%%%%%%%%%%%%%%
\begin{figure*}
\resizebox{\hsize}{!}{\includegraphics[bb=50 70 450 520,
  clip]{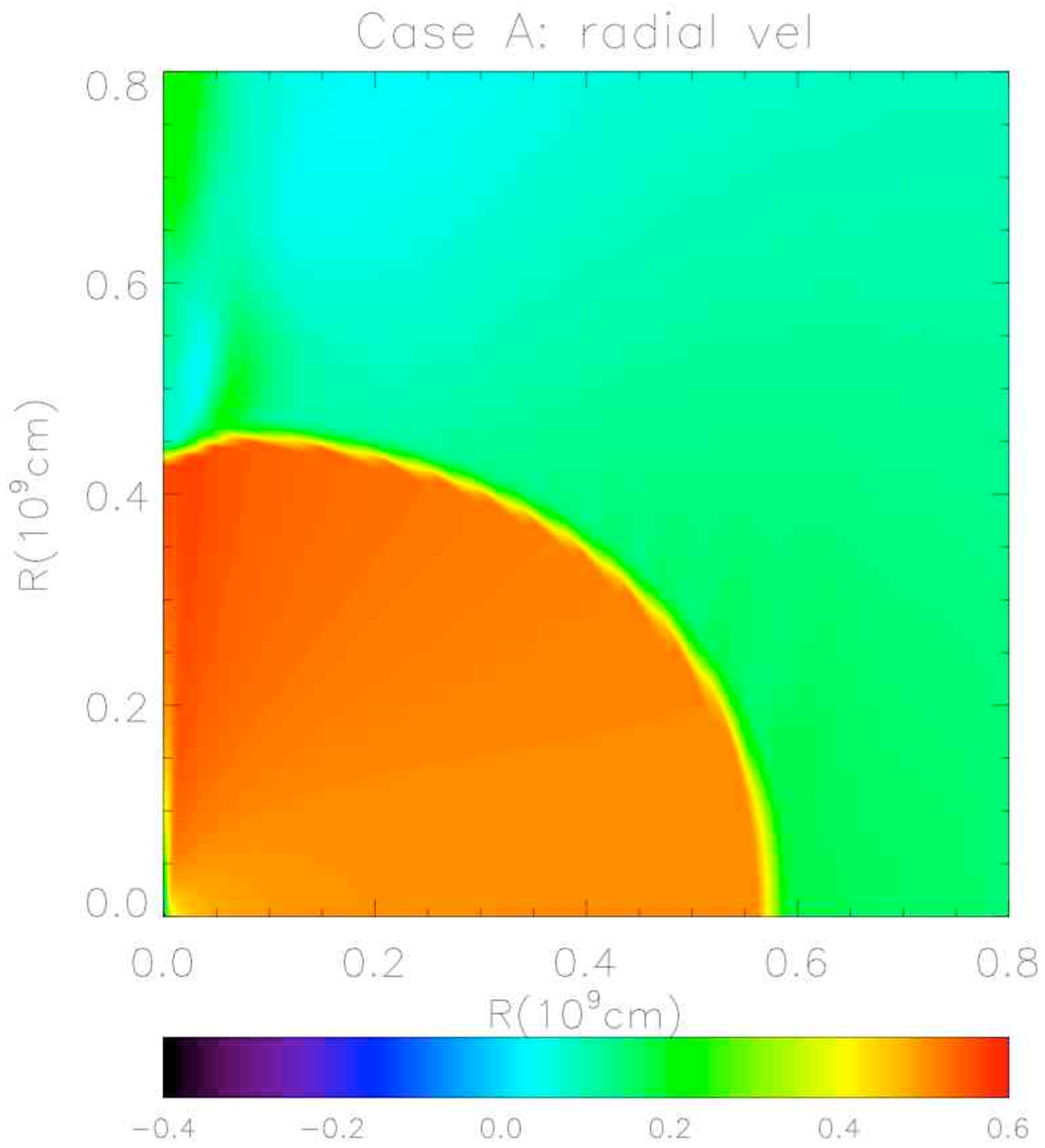},\includegraphics[bb=50 70 450 520,
  clip]{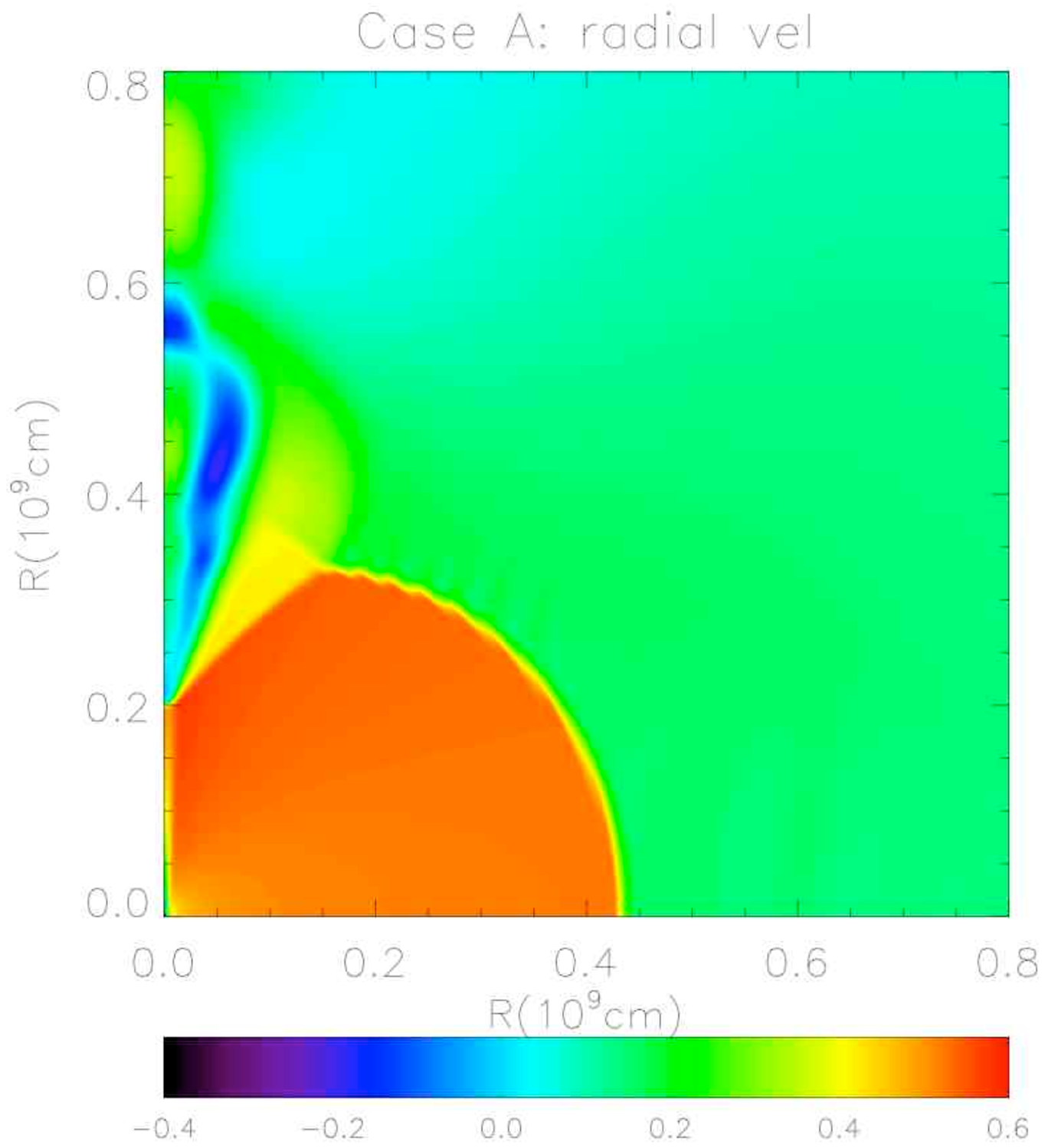},\includegraphics[bb=50 70 450 520, clip]{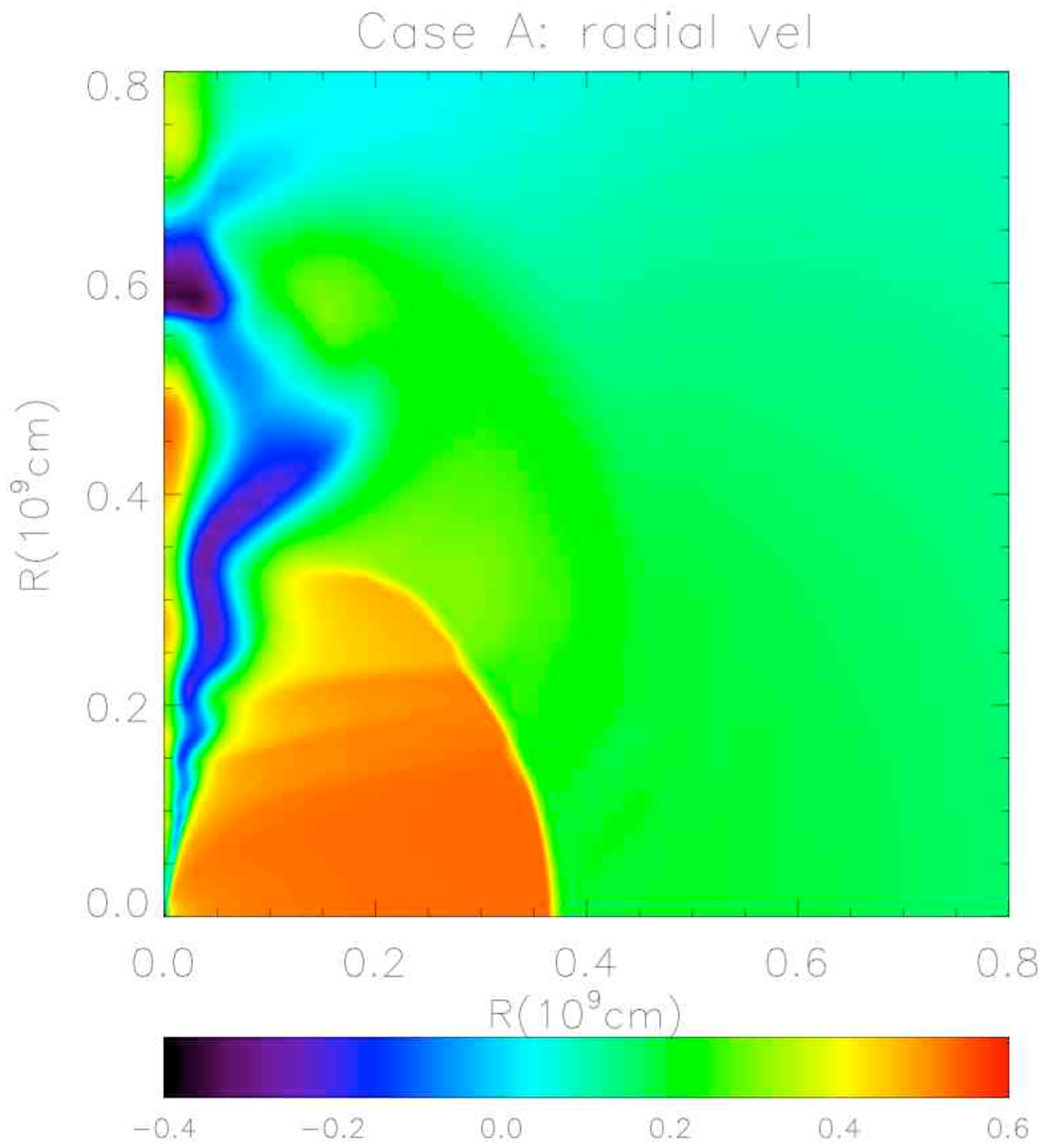}}
%\resizebox{\hsize}{!}{\includegraphics{fig2b.ps}}
%\resizebox{\hsize}{!}{\includegraphics{fig3.ps}}
\caption{Evolution of the radial velocity (in units of c) in a
  magnetized bubble inflated by a magnetar with $B = 10^{15} {\rm G }$
  and $P = 1$ms inside a 35 $M_\odot$ star (Case A), about 2 seconds
  after core bounce (1 second after the start of the simulation).  The
  time difference between the left and right panels is $\simeq 0.15$
  sec.  During this time, the termination shock collapses to small
  radii, inside the fast surface, near the pole, but it remains well
  outside the fast surface at the equator; the fast surface is at
  $\simeq 6 \, 10^6$ cm at the equator and $\simeq 1.2 \, 10^8$ cm at
  the pole. Once the termination shock becomes significantly curved
  near the pole, turbulent eddies are created, as can be identified by
  the negative radial velocity.}
\label{fig:cavit}
\end{figure*}
%%%%%%%%%%%%%%%%%%%%%%%%%%%%%%%%%%%%%%%%%%%%%%%%%%%%%%%%%%%%%%%%%%%%%%%%%%%%

Relativistic MHD codes can have numerical difficulties at sufficiently
large $\sigma$ and/or $B^2/\rho$.  In case C, we found that we were
not able to lower the mass loss rate, and thus increase $\sigma$, at
late times, but had to artificially limit $\sigma$ to be $\le 12$;
this is true even for the free wind.  Despite the fact that the
magnetization is lower than in cases A and B, case C was less
numerically stable. We suspect that this is because the slower
rotation rate in case C leads to a smaller mass loss-rate and thus
lower densities. This affects the stability of the free wind, in which
$B^2/\rho$ grows sufficiently large to cause numerical difficulties at
large distances.  In addition, in Case C there are more complex
interactions between the magnetar wind and the supernova ejecta (see
\S \ref{sec:casec}): the compression of the magnetic field at small
radii leads to the formation of strong current sheets, that the code
fails to handle properly. This happens only at late times, and is due
to the compression of the magnetic field close to the Alfv\'enic
surface at $\sim 10^7$ cm, much smaller than the typical size of the
MWN (see Fig. \ref{fig:zoomC}, discussed below).

\section{Results}
\label{sec:res}
We first describe the results of cases A and B, which are
qualitatively similar.  Case C has a significantly lower spindown
power, and has a somewhat different evolution; this will be discussed
separately at the end.

\subsection{Cases A \& B: High Spindown Power}

We first provide an overview of the evolution of the system as a function
of time.  The density in the initial cavity left by the expanding SN
shock is small enough that the wind from the proto-magnetar can easily
sweep through the cavity in a time $\sim R_{cavity}/V_{wind}\sim 0.1$
sec. As soon as the wind impacts the high density ejecta it is forced
to slow down to a speed of order the SN shock velocity $\sim 0.03c$,
in a strong termination shock (TS). This shock leads to the formation
of a hot magnetized subsonic bubble (the MWN) whose evolution depends
on the magnetization of the wind and on the spindown power of the
proto-neutron star.

In a 1D monopolar geometry it is well-known that the TS can only exist
for a long time at significant distances from the neutron star in the
limit of a weakly magnetized wind (known as the $\sigma$ limit in
PWNe; see \citealt{ken84}). For $\sigma > V_{MWN}/c$ (where $V_{MWN}$
is the expansion speed of the MWN, set initially by the SN shock
velocity in our problem) the shock collapses towards the neutron star
on a timescale of order a sound crossing time, due to the compression
of the toroidal magnetic field.  We have carried out 1D relativistic
MHD simulations of magnetar spindown inside a star and have verified
that for proto-magnetars with rotation rates and magnetic field
strengths comparable to those considered here (see
Table~\ref{table:1}), the TS indeed collapses back down to the neutron
star after $\sim 0.03$ sec.  The resulting causal contact between the
central neutron star and the surrounding progenitor causes the neutron
star to spin-down significantly faster than is predicted by free wind
calculations (e.g., \citealt{met07}).

The evolution is, however, significantly different in the
multidimensional problem considered here. As the wind from the central
neutron star inflates the MWN, two competing effects occur.  On one
hand, magnetic field gets progressively compressed inside the bubble,
causing the TS to recede to smaller radii, as in the 1D case. On the
other hand, the high pressure inside the MWN pushes out against the
denser SN ejecta. For a magnetized wind, the pressure along the
rotation axis is significantly larger than at intermediate latitudes
or the equator, for reasons that we discuss in detail in Papers I and
II.  This leads to a preferential expansion of the MWN in the polar
direction and the formation of an elongated bubble.  As a result, the
$\sigma $ limit is less severe in 2D than in 1D: magnetic field is
allowed to flow from the equatorial region to the polar region in
order to establish the magnetohydrostatic pressure distribution
\citep{beg92}.  For large spindown power (Cases A \& B), the expansion
of the MWN in the polar region is fast enough to relieve the
compression of the magnetic field by allowing an escape channel along
the axis.  As a consequence the TS moves towards the central neutron
star at a slower speed than in 1D. Once the jet emerges from the
central part of the star, the outflow through the polar channel
compensates for the compression of the magnetic field and the TS
stabilizes.

Figure \ref{fig:cavit} shows the radial velocity (in units of c) at
three different times for Case A, all approximately 2 second after
core bounce; the three times shown are within $\simeq 0.15$ sec of
each other.  Initially, the TS is outside the fast surface at all
latitudes and its shape is roughly spherical.  As discussed in
previous papers about PWNe \citep{ldz04} the shape of the TS can be
understood qualitatively in terms of pressure balance between the wind
and the nebula. At early times the wind is moderately relativistic and
is collimated in the polar direction; the ram pressure is thus higher
along the axis than at the equator. The nebula also has a higher axial
pressure due to the compressed toroidal magnetic field
\citep{beg92}. These two effects roughly balance and as a result the
TS is roughly spherical. However, at later times the proto-magnetar
wind becomes progressively more magnetically dominated, due to the
decreasing neutrino flux from the proto-neutron star, and the
associated drop in mass loss rate. The spindown
power thus becomes progressively more equatorial. At the same time,
the compression of the toroidal magnetic field in the MWN tends to
increase the pressure anisotropy in the bubble. As a result the TS
becomes oblate and a cusp forms at the pole, as can be seen at the
later times in Figure~\ref{fig:cavit}. As soon as the polar cusp forms
the inclination of the termination shock causes the formation of
vorticity in the post shock region; this leads to the formation of
large scale eddies along the axis.  Because of the angle between the
wind and the shock, the cusp also strongly reduces the effective ram
pressure exerted by the wind.

At later times, the TS along the axis moves to much smaller radii, inside
the location of the fast surface (which along the axis it at a much
larger distance from the neutron star than on the equator; e.g.,
\citealt{me06}). Once the TS reaches the fast surface it becomes a
weak Alfv\'enic discontinuity and a sub-fast outflow is established in
the polar region which puts the nebula in causal contact with the
central engine (Fig.~\ref{fig:cavit}). However in both cases A and B
the shock in the equatorial region never reaches the location of the
fast surface, for two reasons: first, the TS shock tends to be at
larger radii in the equator because the nebular pressure in the
equator is lower than at the pole while for high $\sigma$ the wind ram
pressure is higher at the equator; and, second, the fast surface is
closer to the central engine in the equator ($\simeq 60$ km at the
equator and $\simeq 10^8$ cm at the pole). The result is a mixed
outflow in which the flow is subfast (but super-Alfv\'enic) in the
polar region but superfast in the equatorial region. We will discuss
the implications of this causal contact for the spindown of the
neutron star in \S \ref{sec:torque}.

%%%%%%%%%%%%%%%%%%%%%%%%%%%%%%%%%%%%%% FIG 2 %%%%%%%%%%%%%%%%%%%%%%%%%%%%%%
\begin{figure*}
\resizebox{\hsize}{!}{\includegraphics[bb=50 70 450 520, clip]{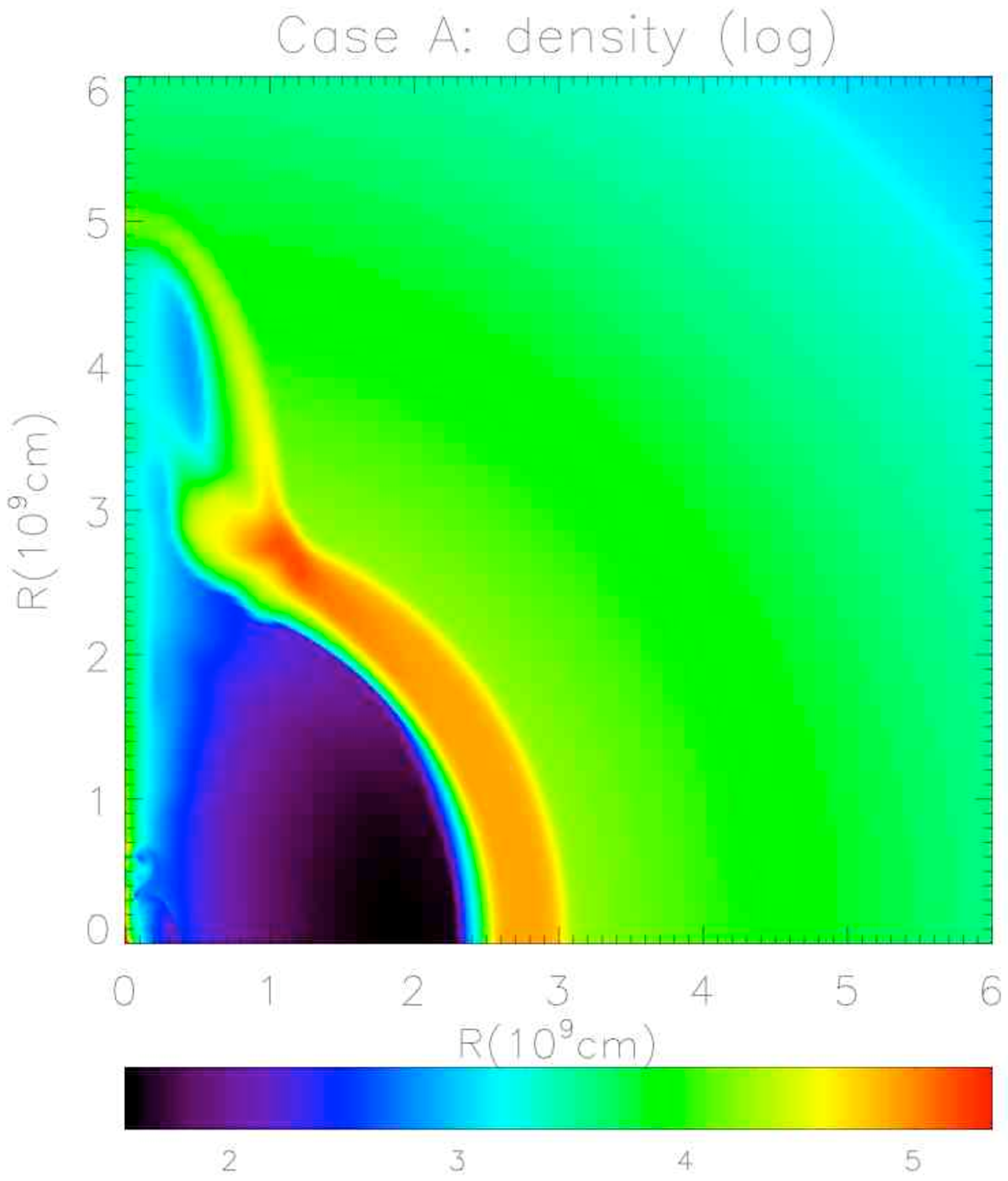}\includegraphics[bb=50 70 450 520, clip]{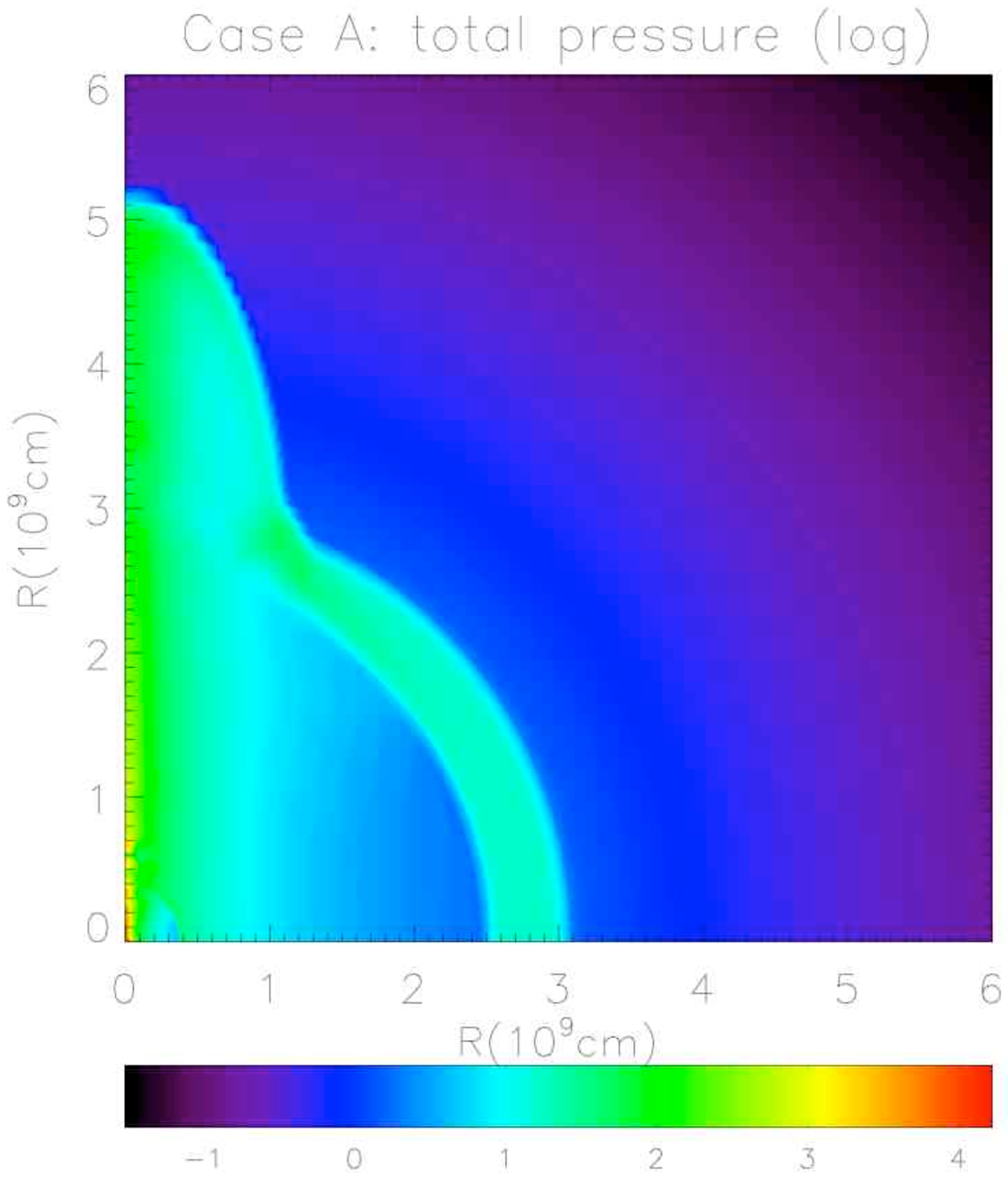}\includegraphics[bb=50 70 450 520, clip]{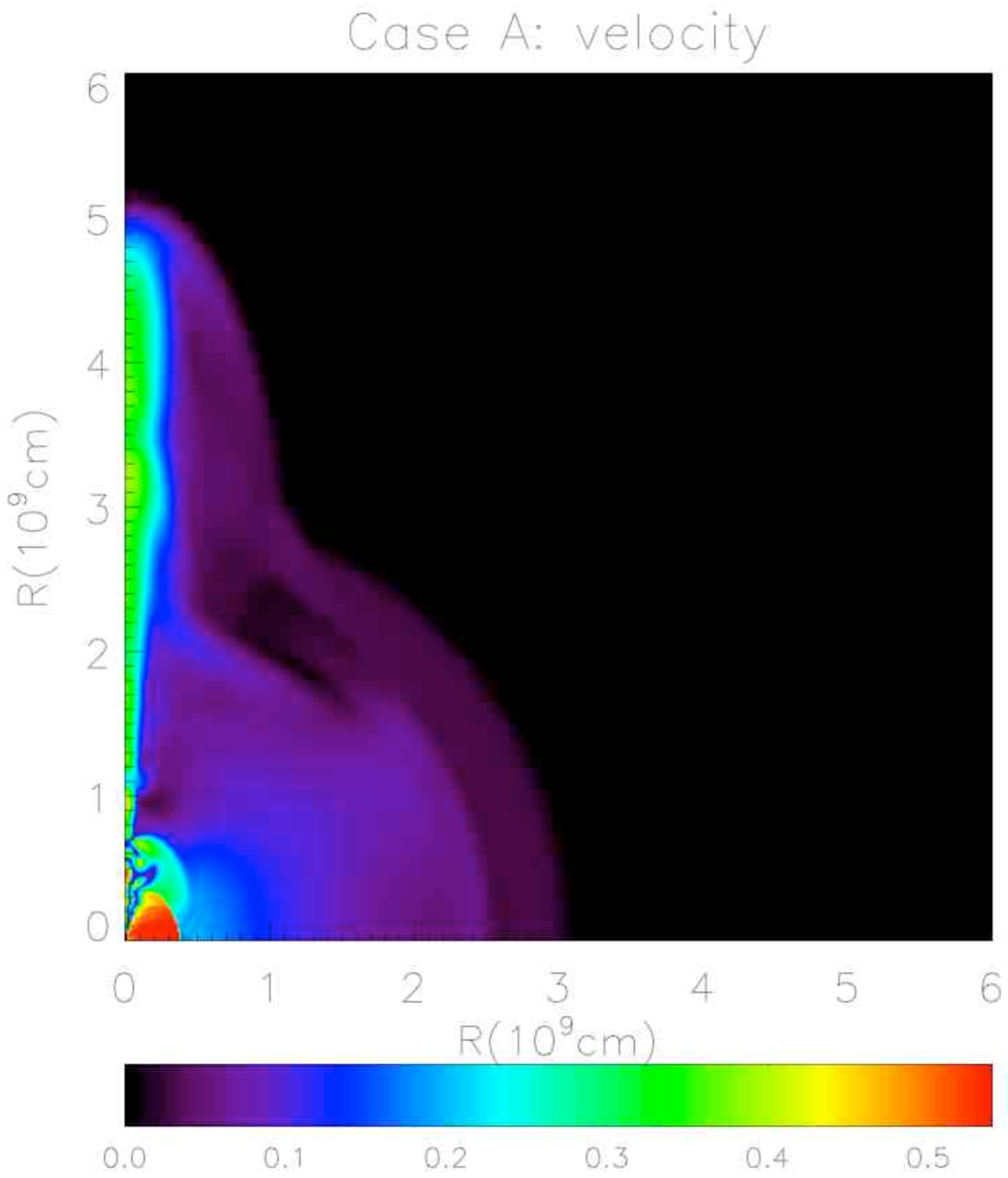}}
\resizebox{\hsize}{!}{\includegraphics[bb=50 70 450 520, clip]{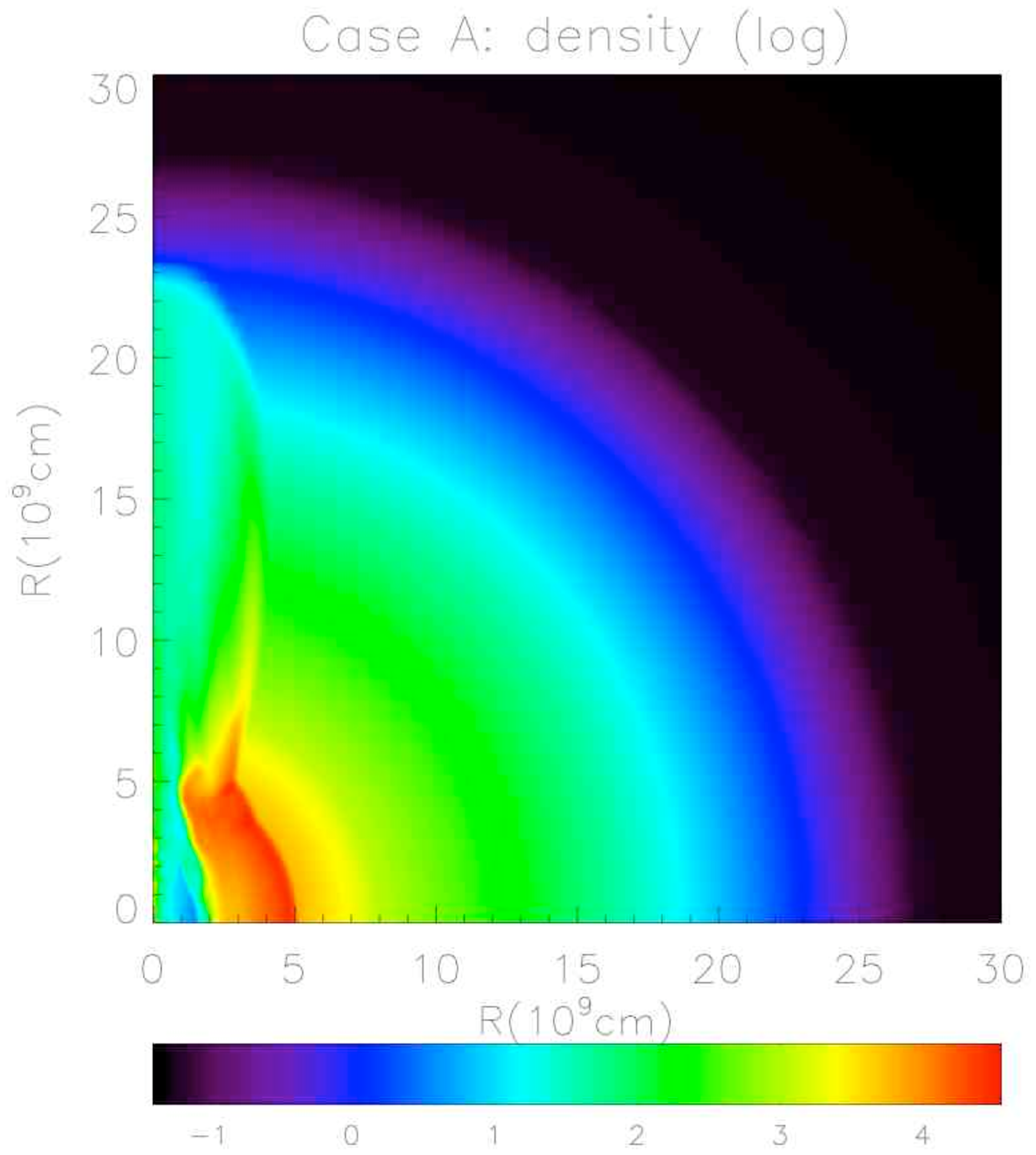}\includegraphics[bb=50 70 450 520, clip]{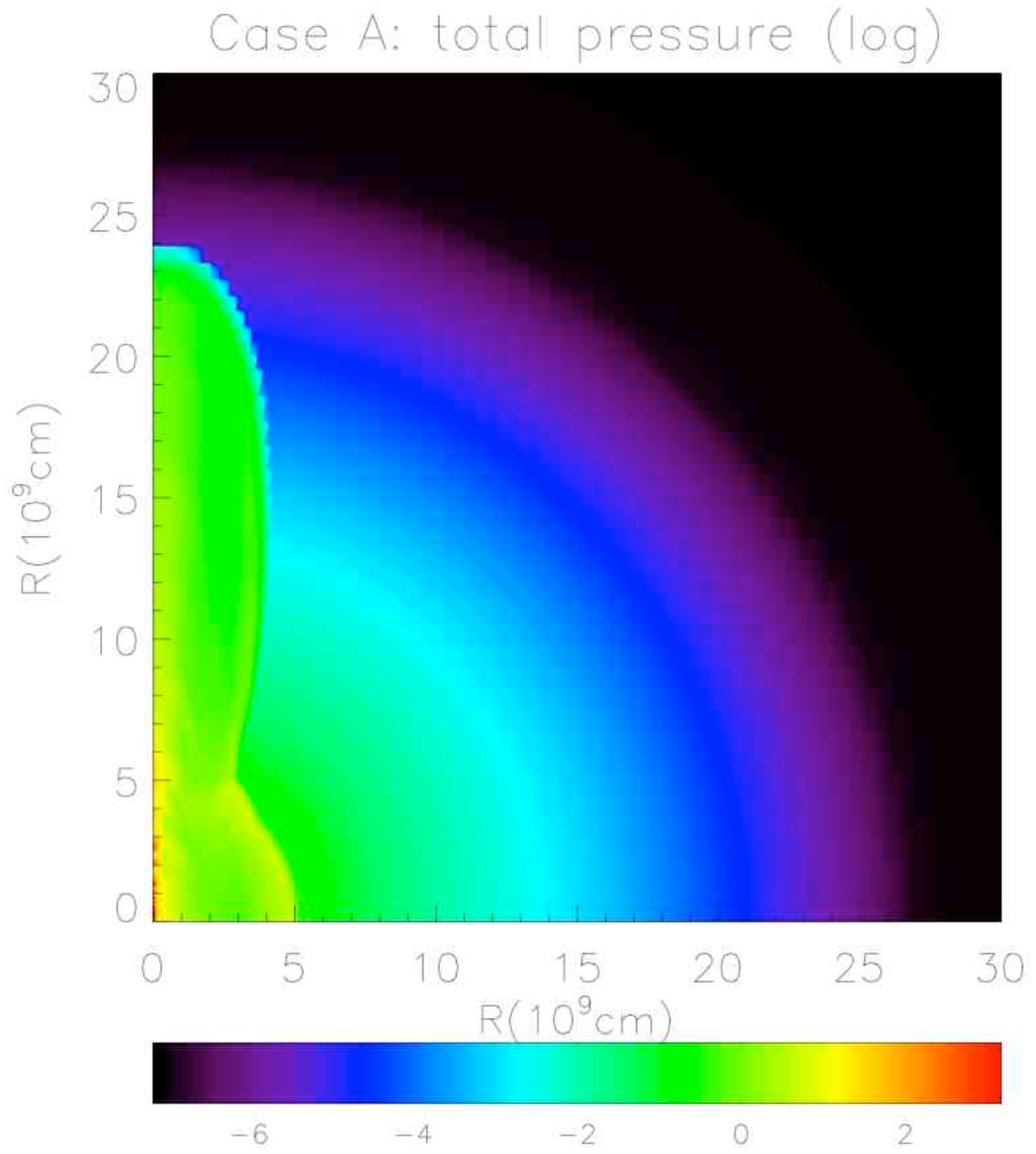}\includegraphics[bb=50 70 450 520, clip]{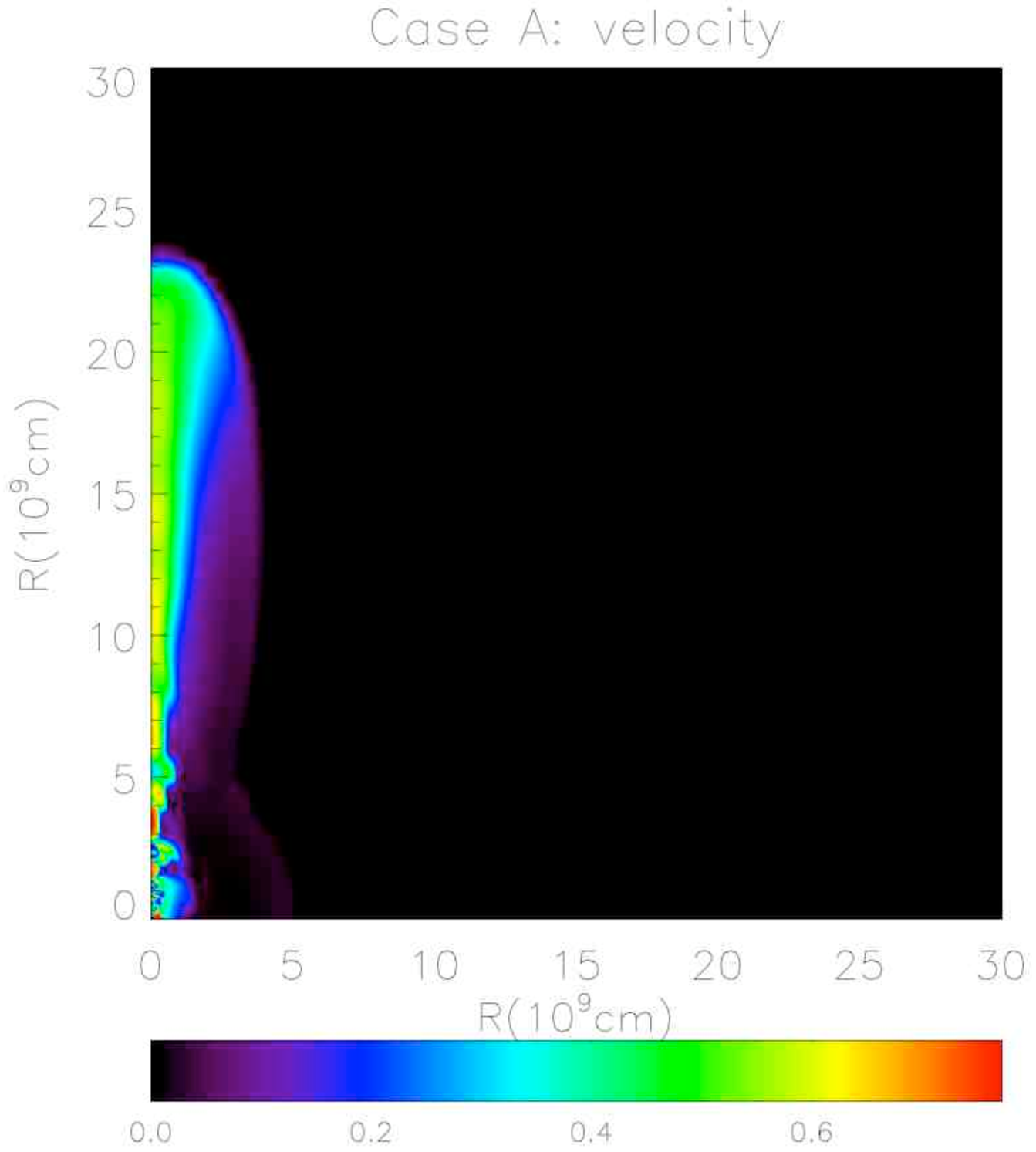}}
\resizebox{\hsize}{!}{\includegraphics[bb=50 70 450 520, clip]{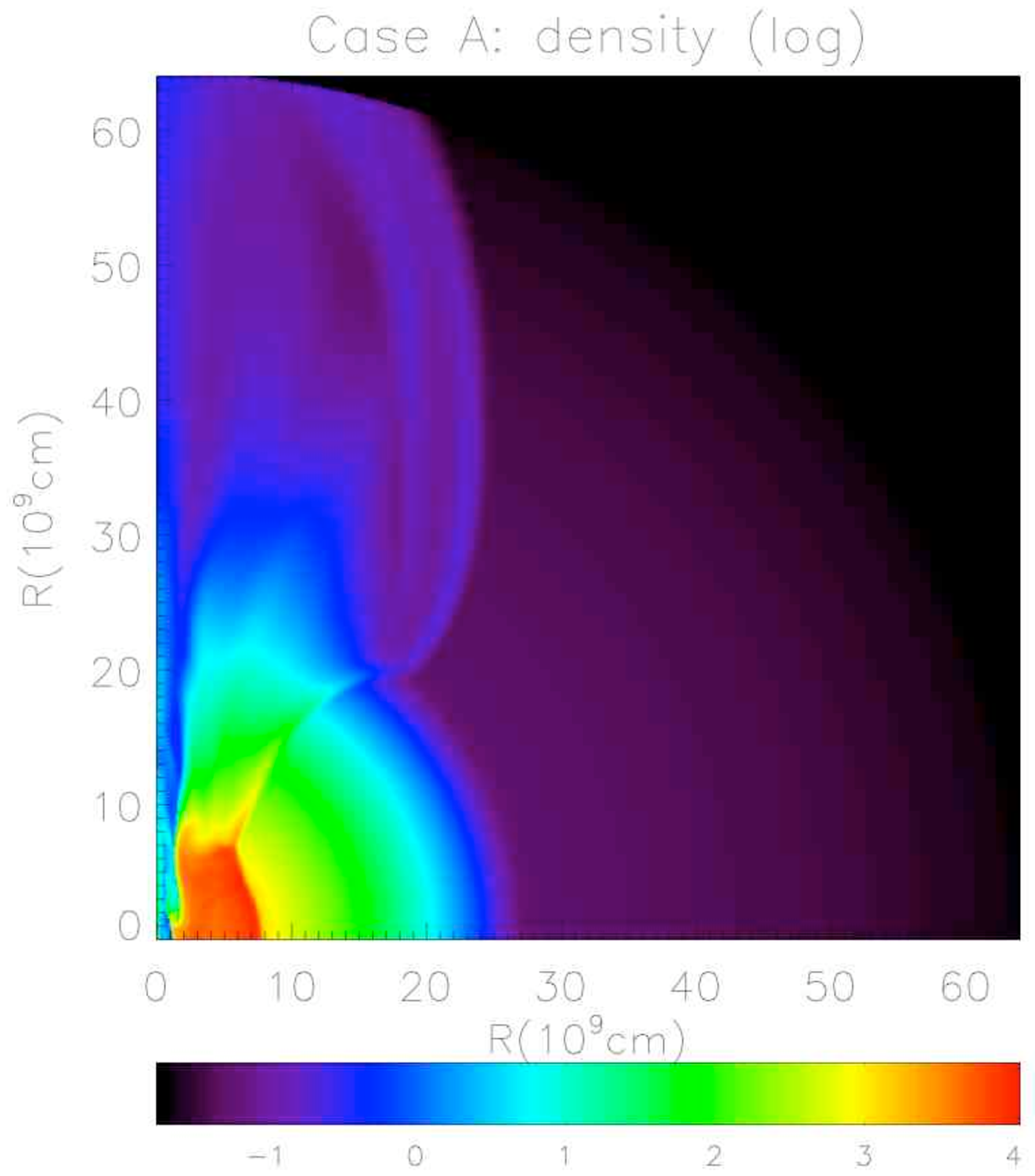}\includegraphics[bb=50 70 450 520, clip]{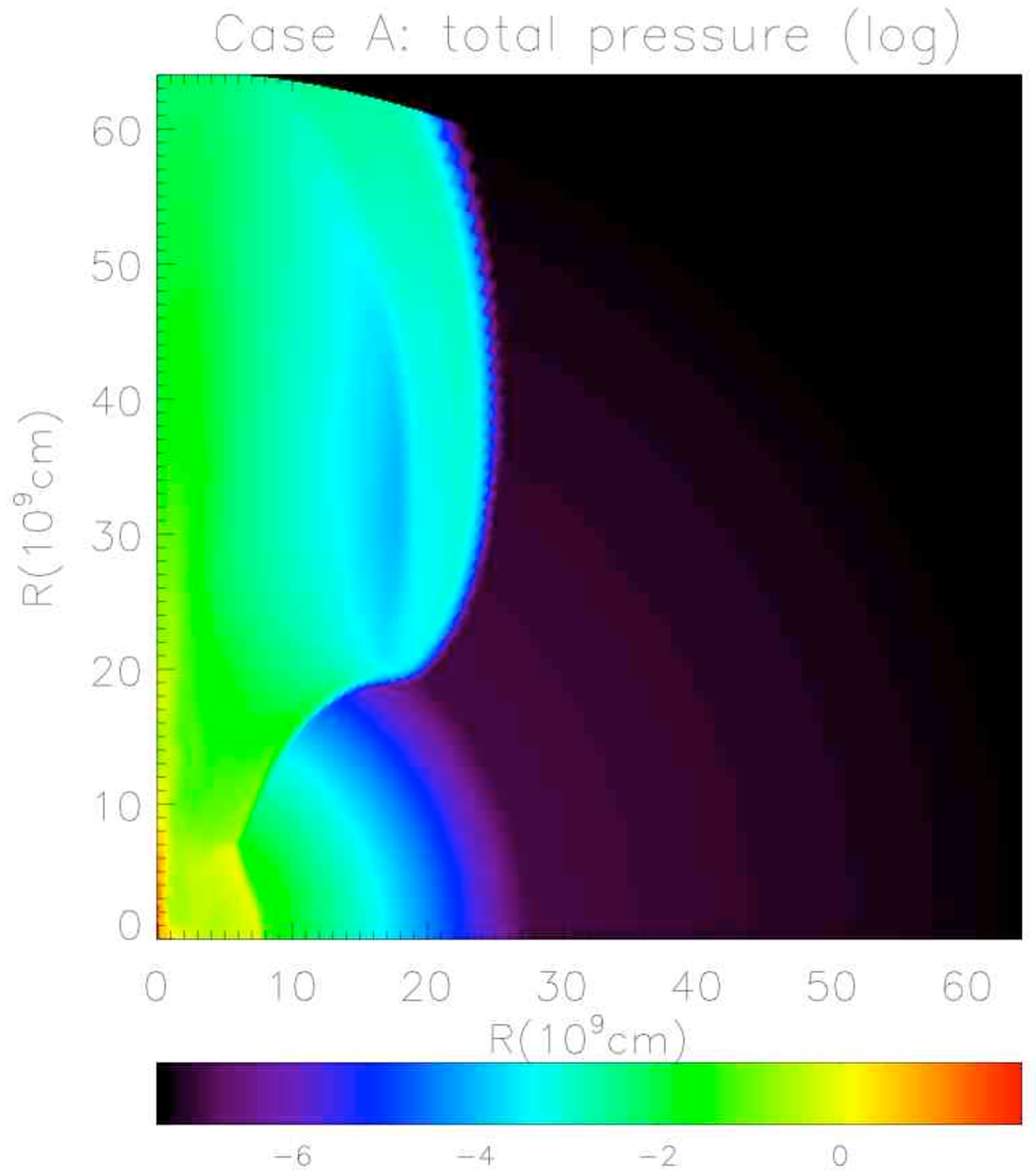}\includegraphics[bb=50 70 450 520, clip]{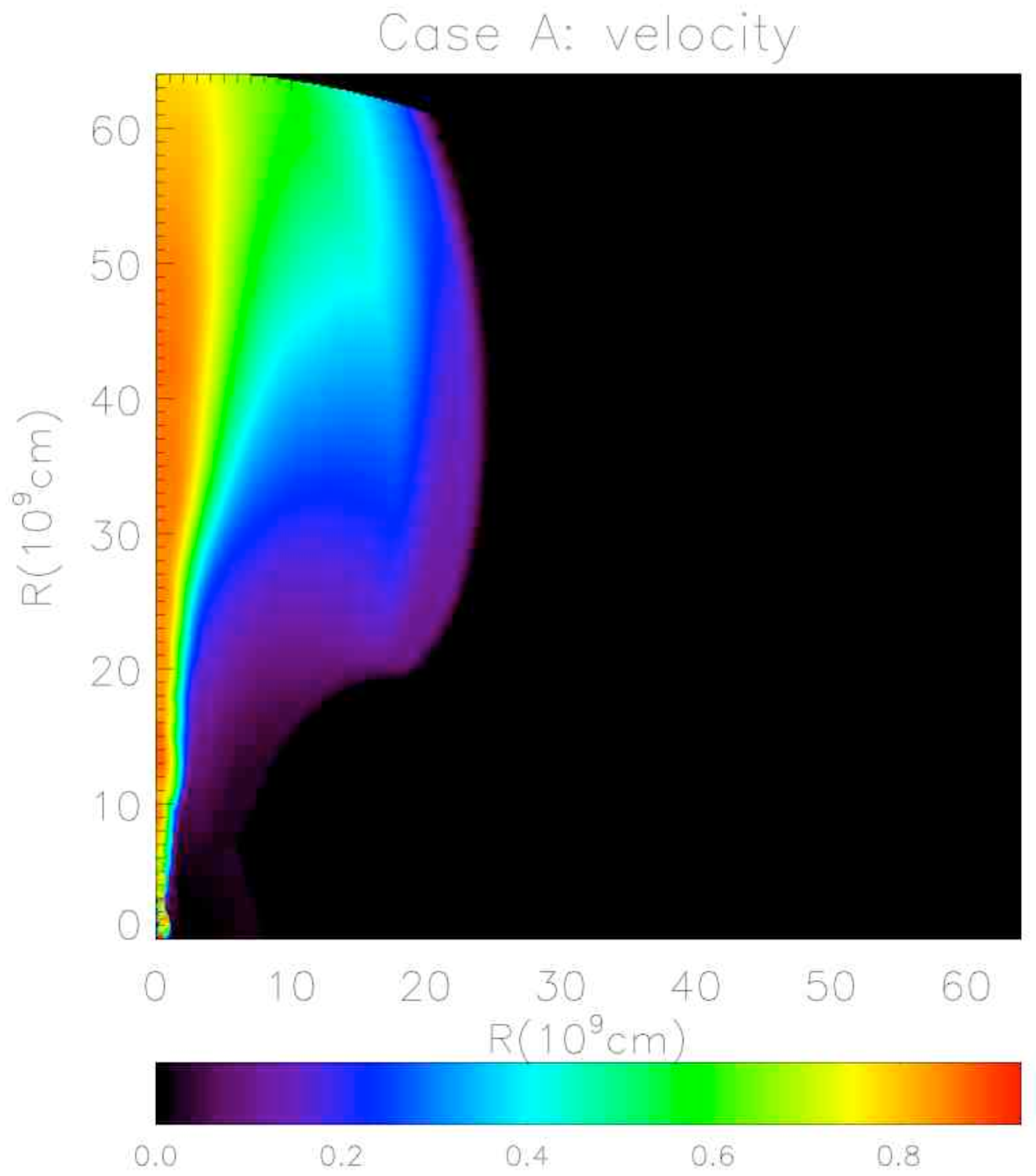}}
\caption{Evolution of the magnetized bubble inflated by a magnetar
  with $B = 10^{15} {\rm G }$ and $P = 1$ms inside a 35 $M_\odot$ star
  (Case A). From left to right, the panels show: log$_{10}$[density
  (g cm$^{-3}$)], log$_{10}$[pressure (erg cm$^{-3}\, c^2$)] and
  velocity (in units of $c$). From top to bottom, the snapshots are
  $2,4,$ \& $7$ seconds after core bounce. The radius of the
  progenitor star is $2.5 \, 10^{10}$ cm.  By $t \sim 5$ sec (middle
  panel) the jet has escaped the progenitor star.}
\label{fig:caseA}
\end{figure*}
%%%%%%%%%%%%%%%%%%%%%%%%%%%%%%%%%%%%%%%%%%%%%%%%%%%%%%%%%%%%%%%%%%%%%%%%%%%%

%%%%%%%%%%%%%%%%%%%%%%%%%%%%%%%%%%%%%% FIG 3 %%%%%%%%%%%%%%%%%%%%%%%%%%%%%%
\begin{figure*}
\resizebox{\hsize}{!}{\includegraphics[bb=50 70 450 520, clip]{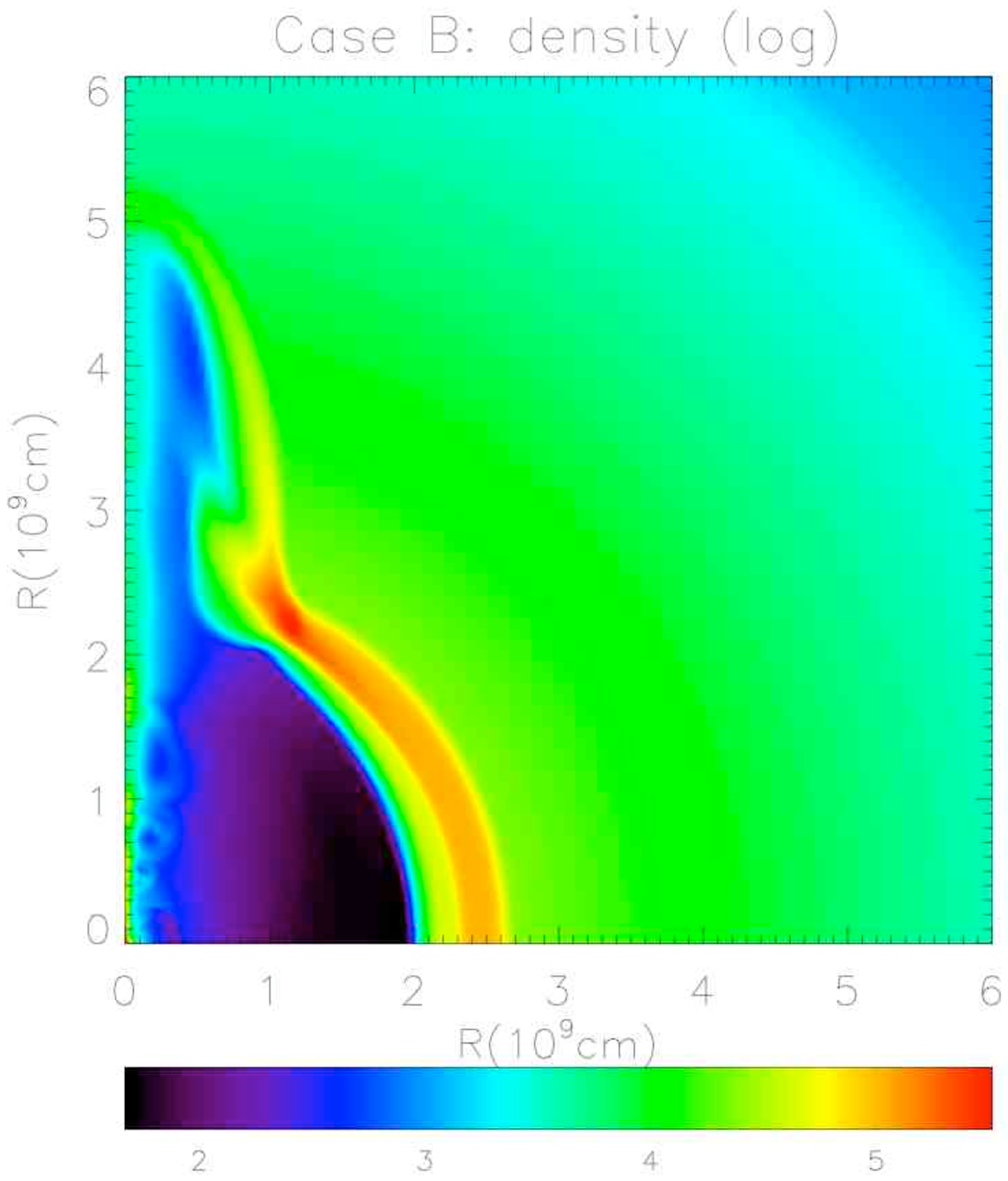}\includegraphics[bb=50 70 450 520, clip]{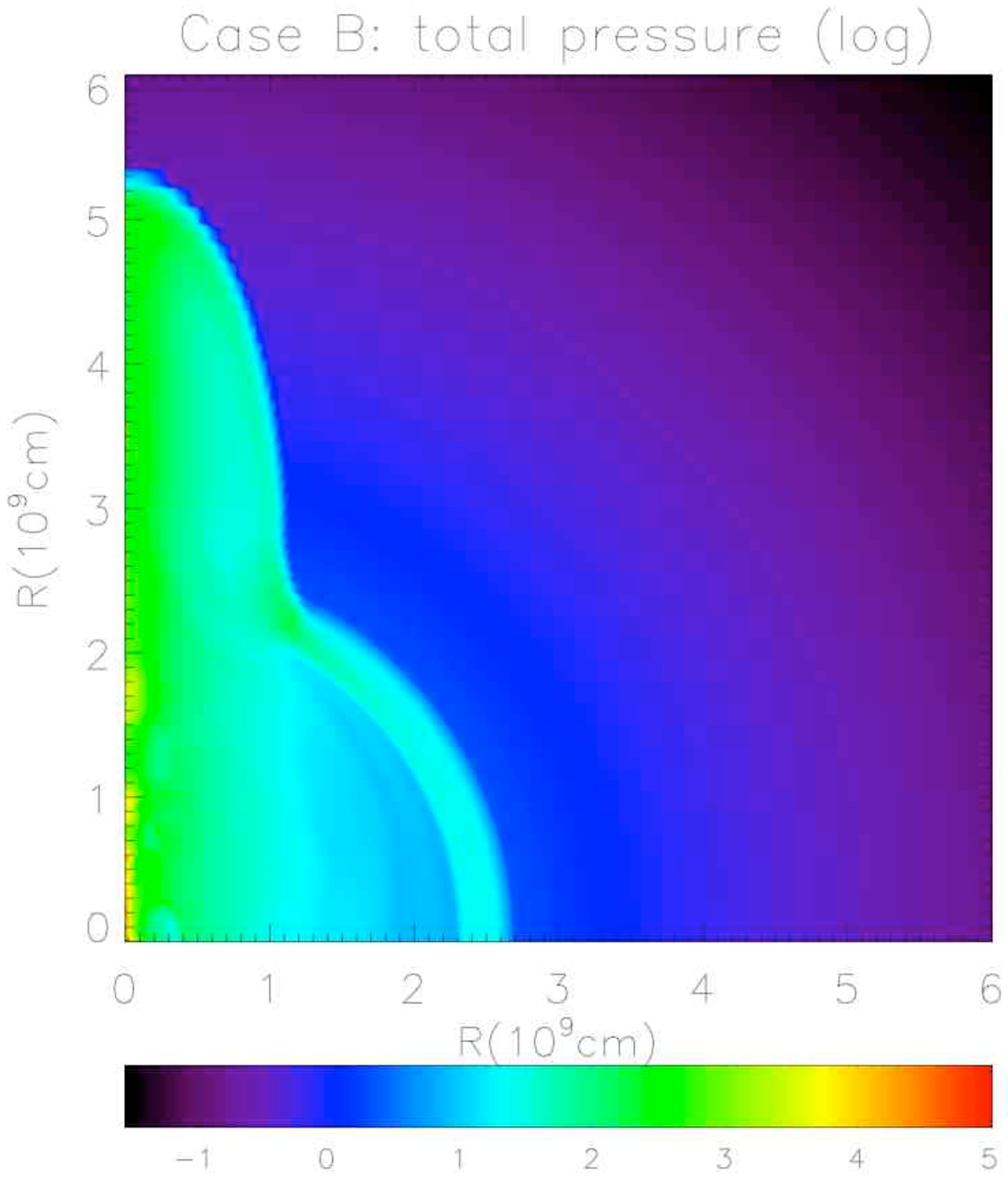}\includegraphics[bb=50 70 450 520, clip]{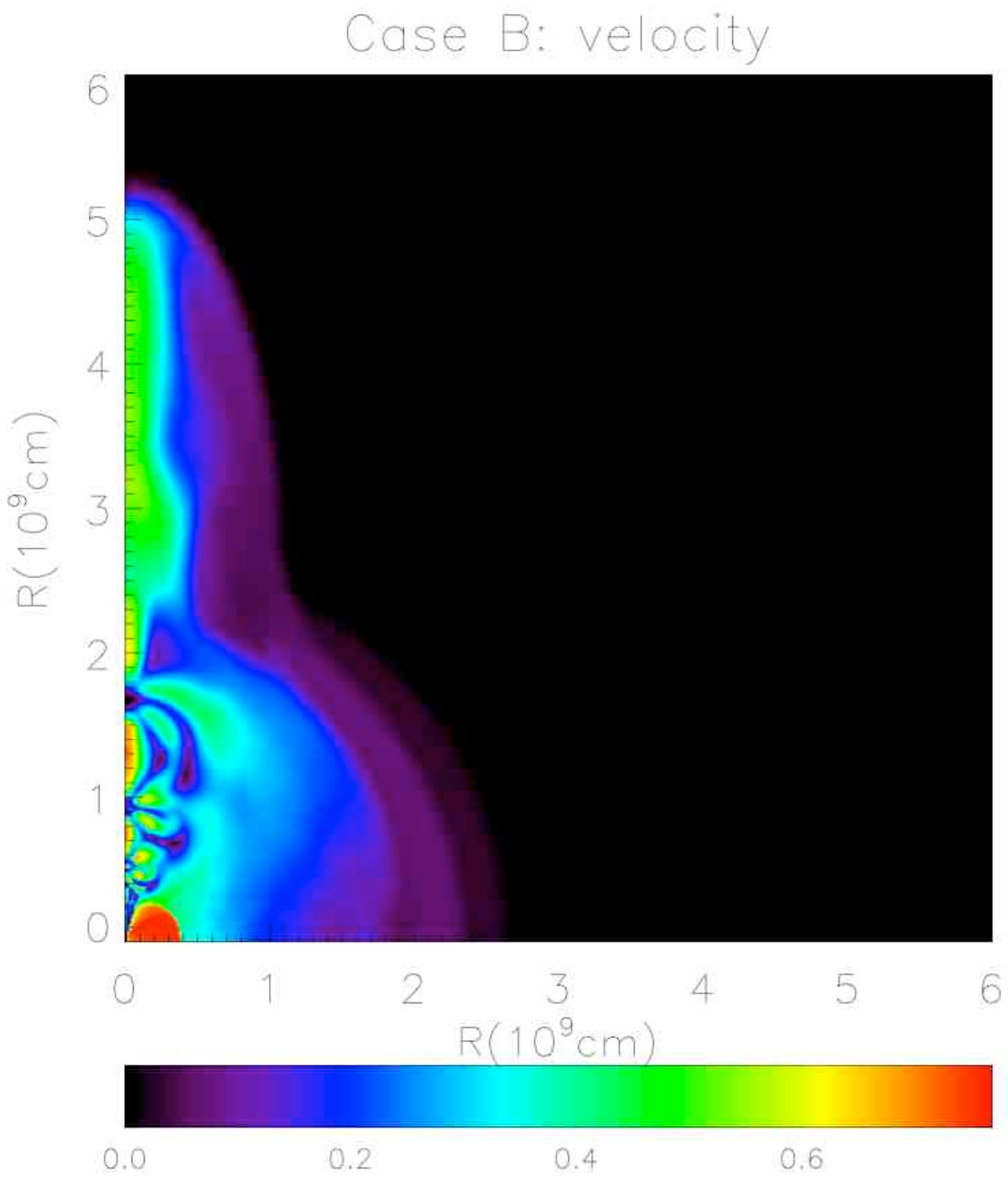}}
\resizebox{\hsize}{!}{\includegraphics[bb=50 70 450 520, clip]{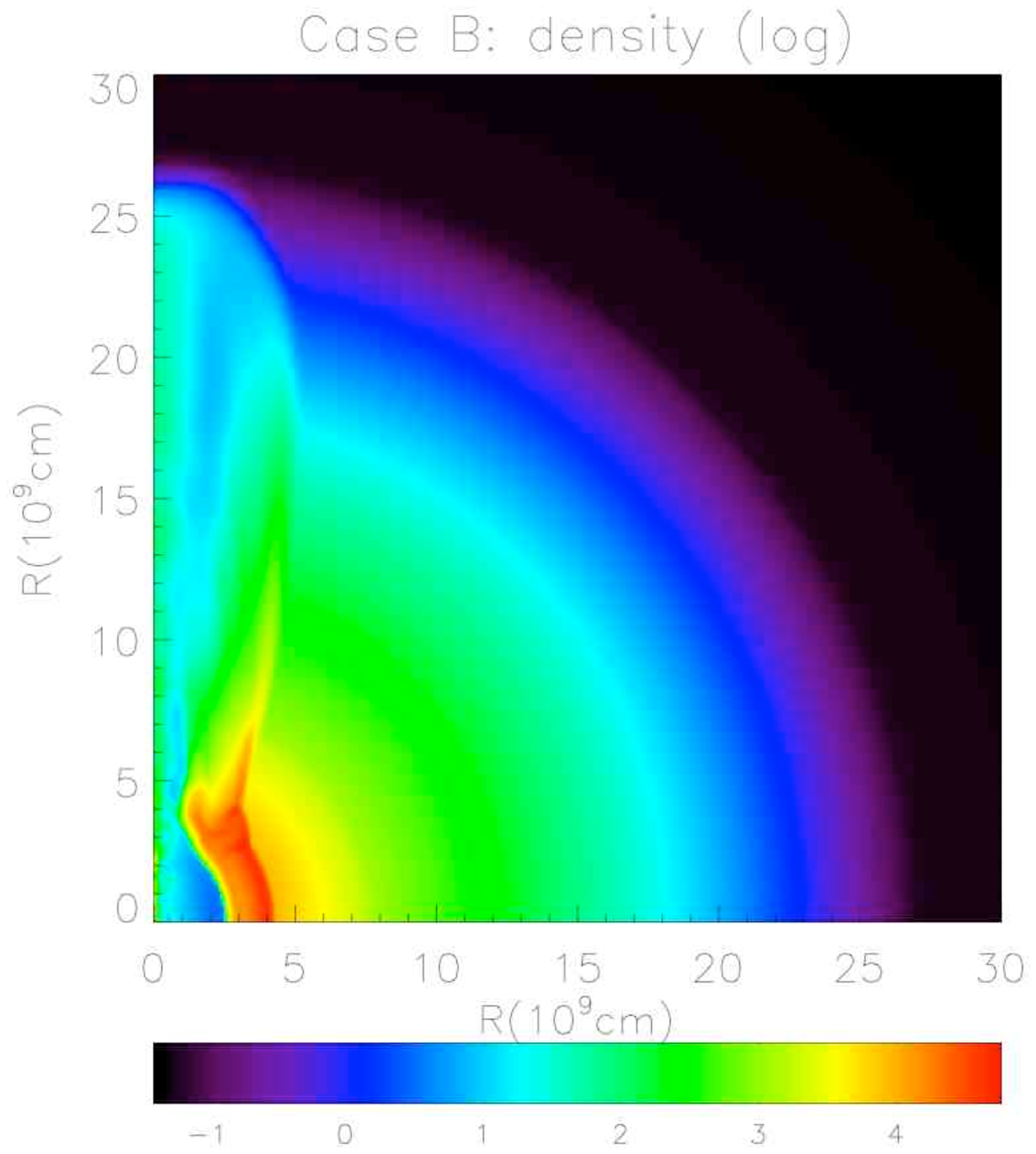}\includegraphics[bb=50 70 450 520, clip]{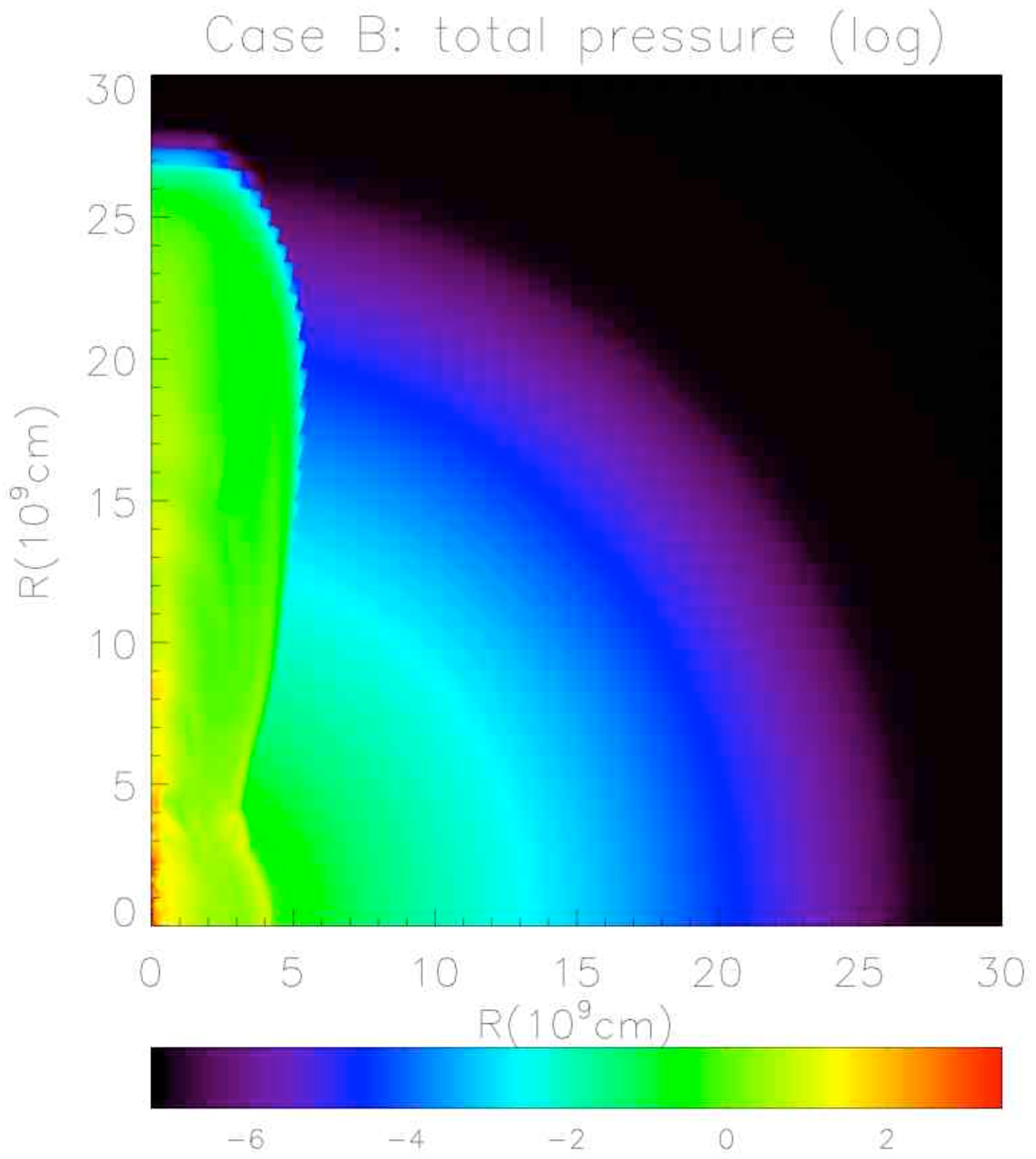}\includegraphics[bb=50 70 450 520, clip]{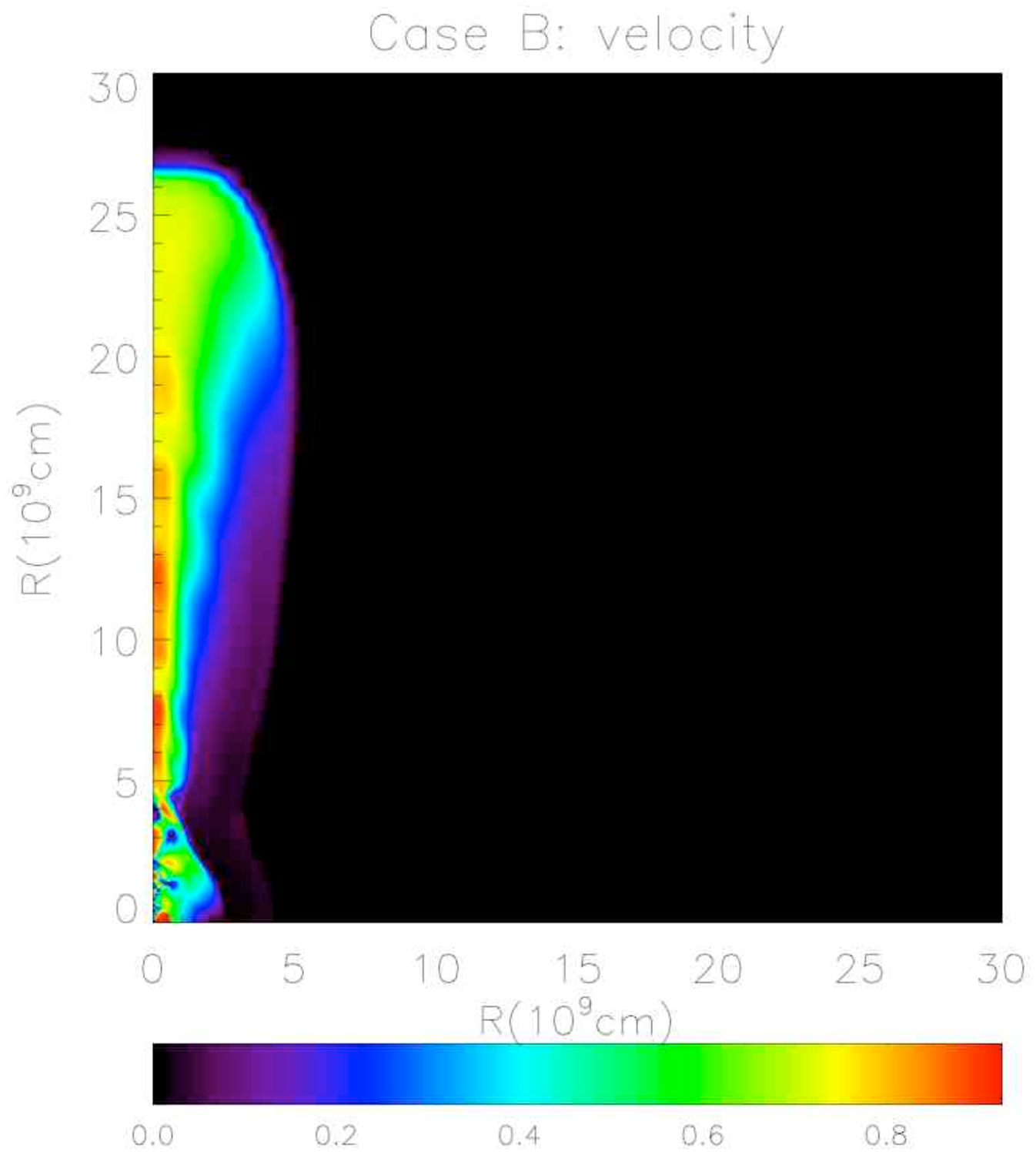}}
\resizebox{\hsize}{!}{\includegraphics[bb=50 70 450 520, clip]{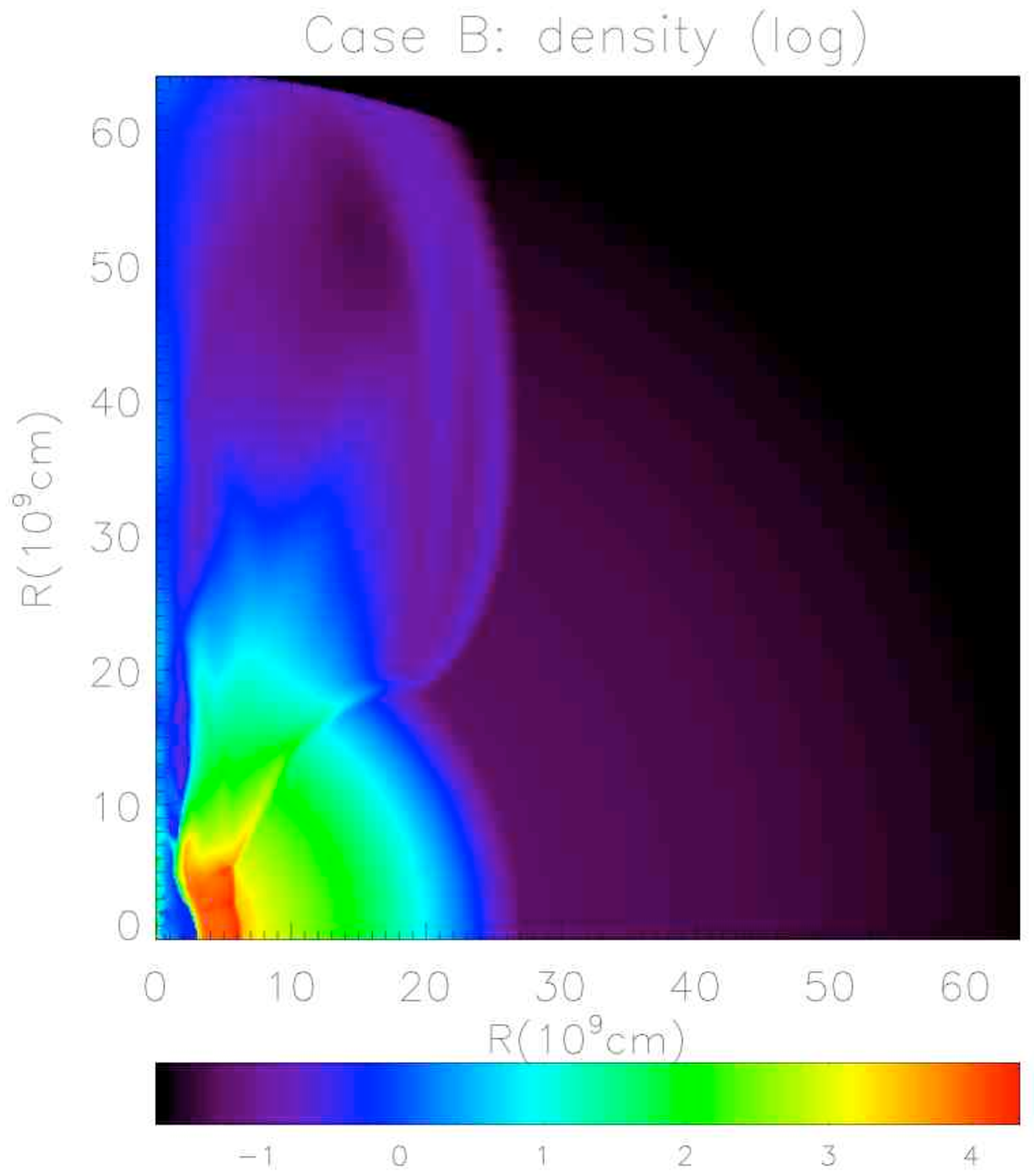}\includegraphics[bb=50 70 450 520, clip]{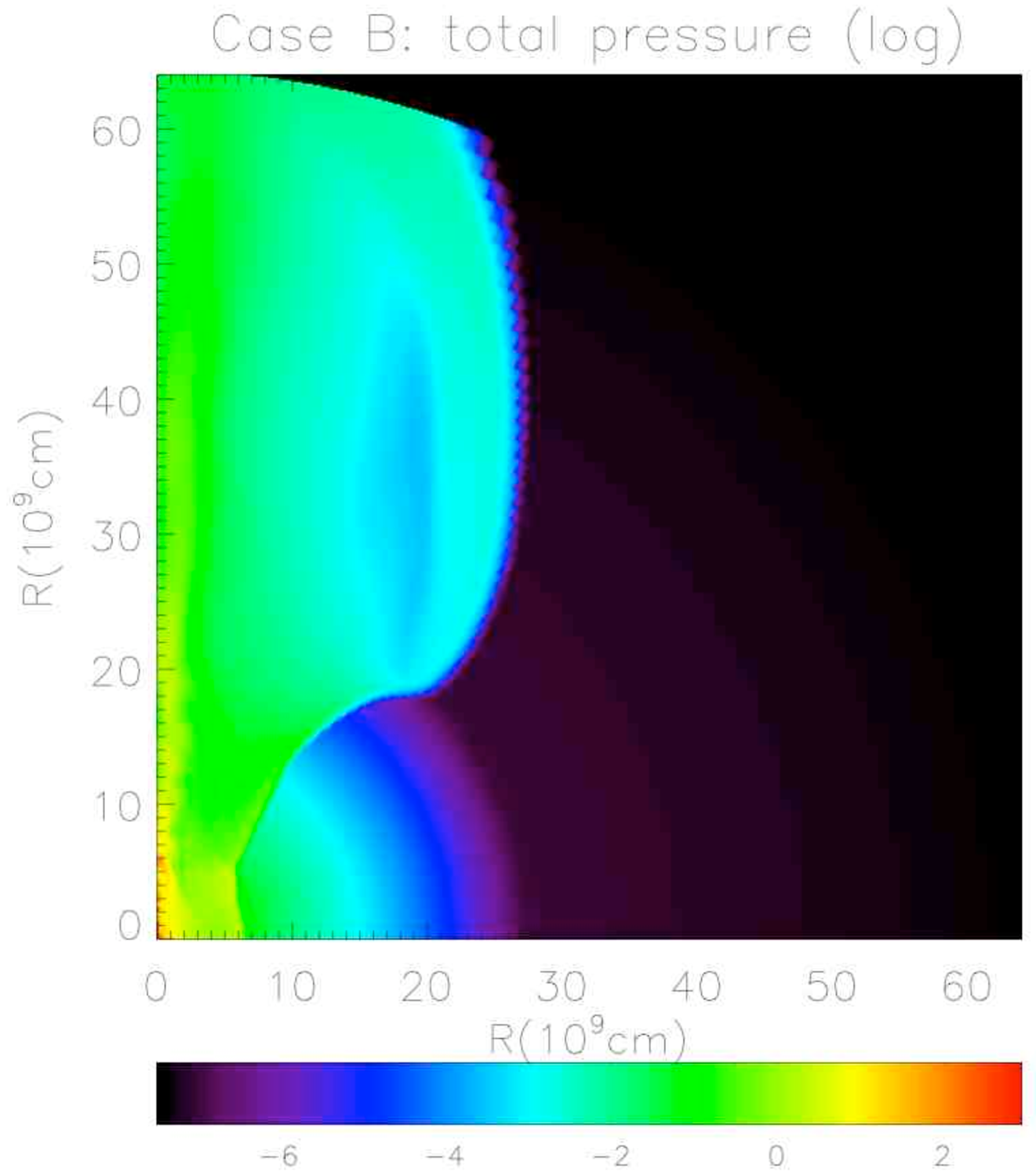}\includegraphics[bb=50 70 450 520, clip]{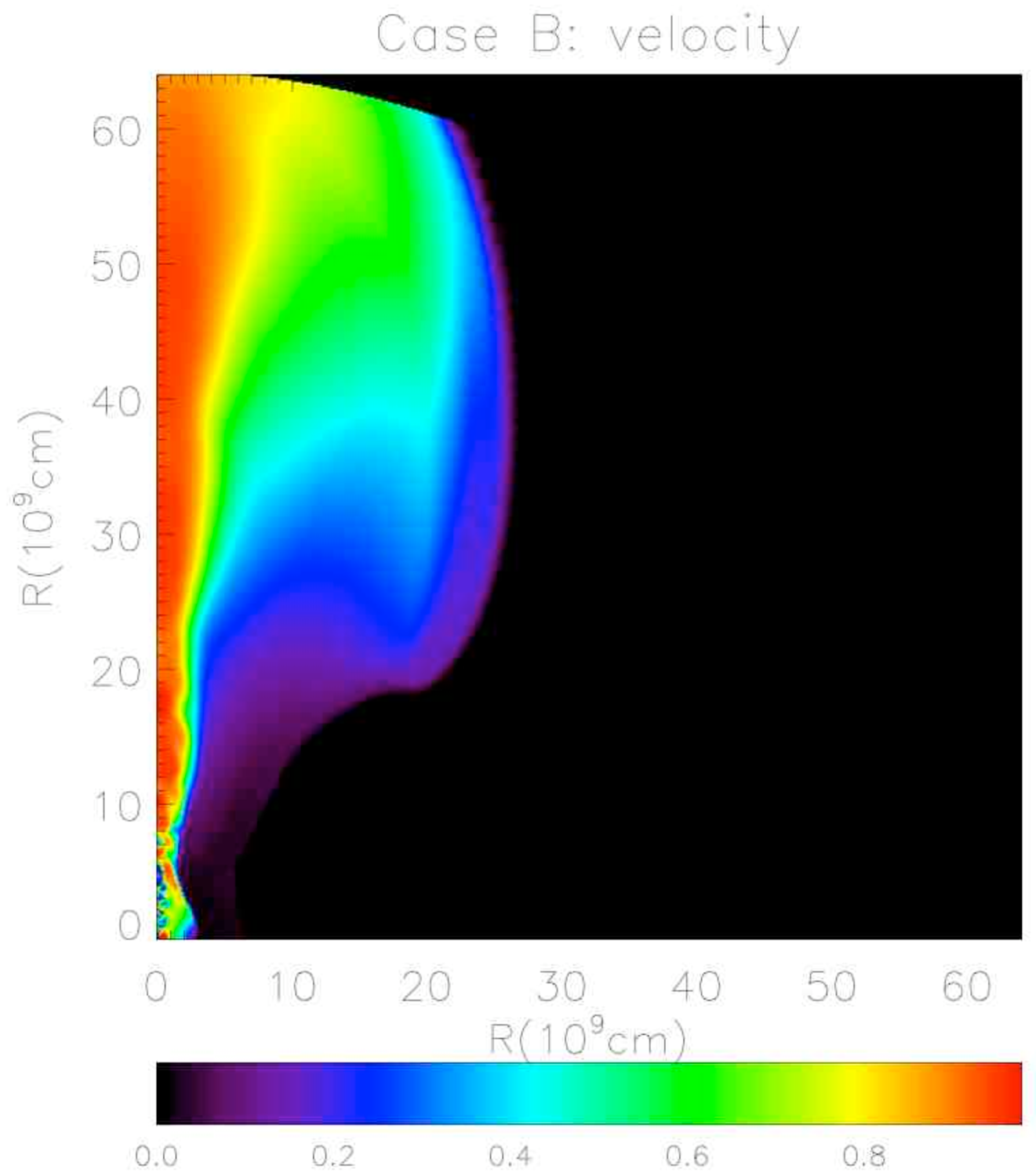}}
\caption{Evolution of the magnetized bubble inflated by a magnetar
  with $B = 3 \times 10^{15} {\rm G }$ and $P = 1$ms inside a 35
  $M_\odot$ star (Case B). From left to right, the panels show:
  log$_{10}$[density (g cm$^{-3}$)], log$_{10}$[pressure (erg
  cm$^{-3}\, c^2$)] and velocity (in units of $c$). From top to
  bottom, the snapshots are $1.6,3,$ \& $5$ seconds after core
  bounce. The radius of the progenitor star is $2.5 \, 10^{10}$ cm.
  By $t \sim 3$ sec (middle panel) the jet has escaped the progenitor
  star.}
\label{fig:caseB}
\end{figure*}
%%%%%%%%%%%%%%%%%%%%%%%%%%%%%%%%%%%%%%%%%%%%%%%%%%%%%%%%%%%%%%%%%%%%%%%%%%%%

In both cases A and B the expansion in the polar region is fast enough
that even at relatively early times, the MWN has already created a
jet-like feature along the rotation axis that is expanding at speeds
that are a reasonable fraction of the speed of light. Figures
\ref{fig:caseA} \& \ref{fig:caseB} show the later time evolution of
the density, pressure, and velocity, for cases A and B,
respectively. The polar jet expands much more rapidly than the SN
shock and by $\sim 3-5$ sec the jet is outside the surface of the
progenitor star.  Despite some quantitative differences, the late-time
evolution of the MWN is quite similar to that found in Paper II, in
which we injected a high $\gamma$, low $\sigma$ wind.

\begin{figure*}
\resizebox{\hsize}{!}{\includegraphics[bb= 93 290 463
  715, clip]{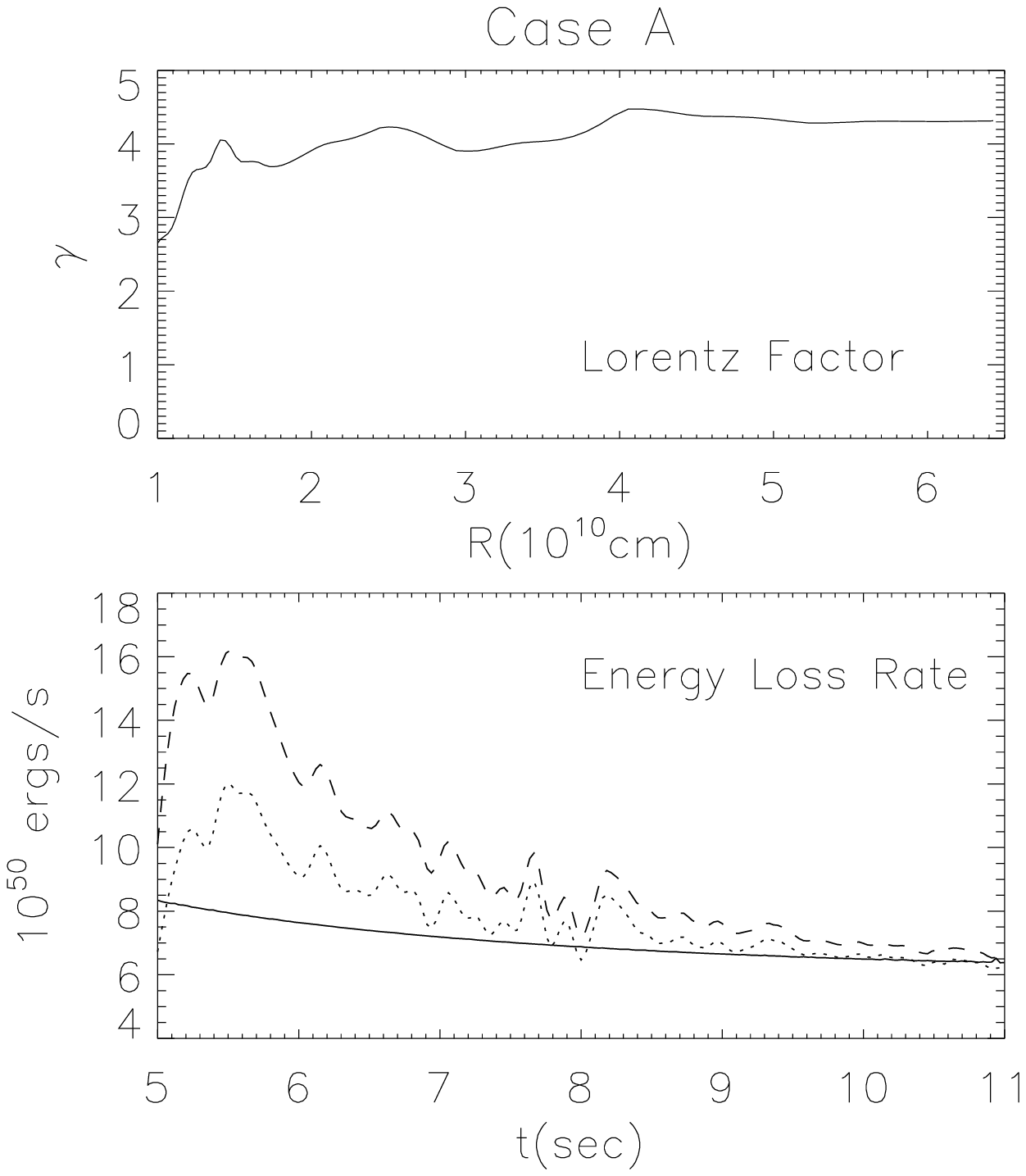}\includegraphics[bb= 77 578 463 996, clip]{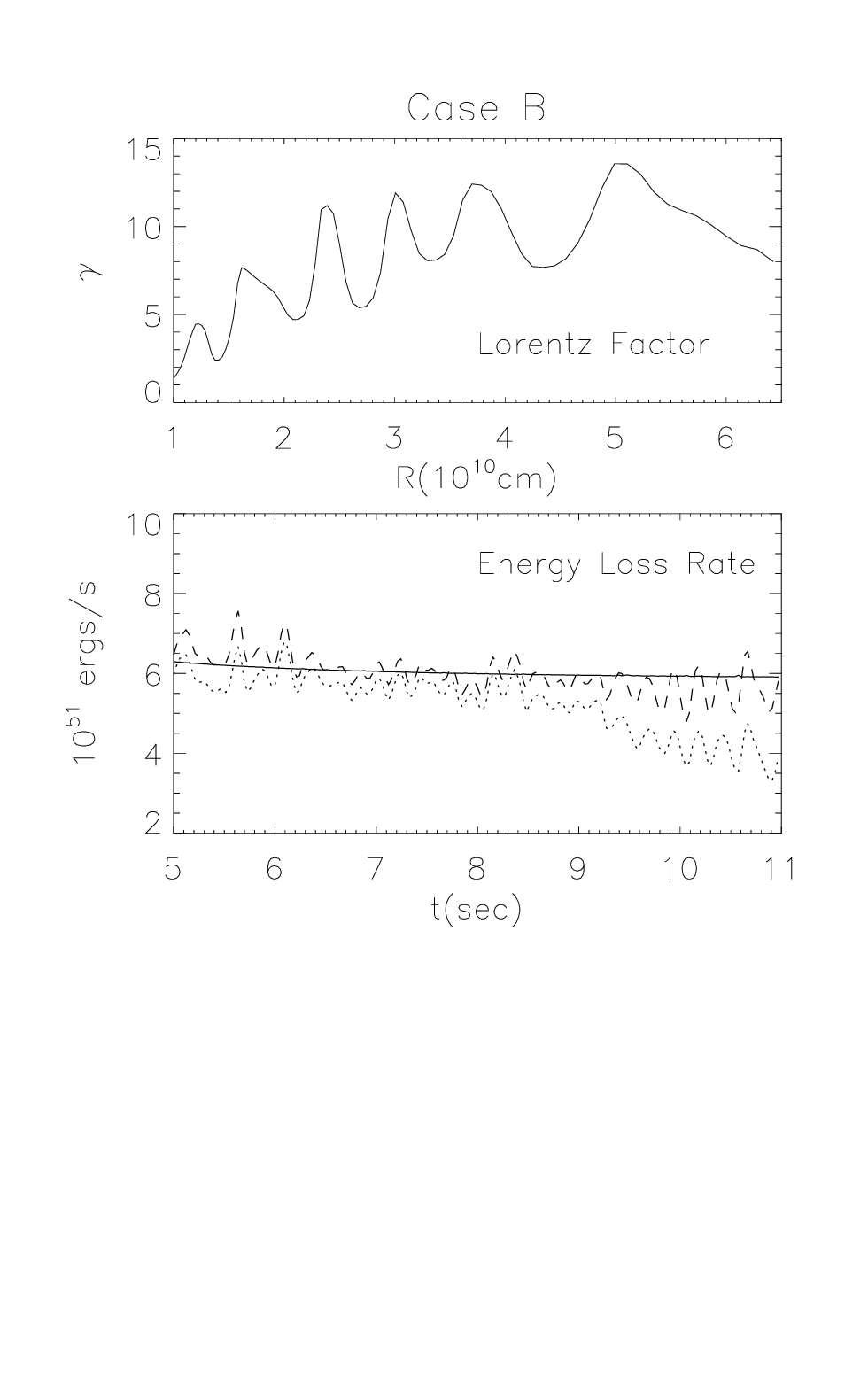}}
\caption{Outflow and spindown properties for Cases A (left) and B
  (right). {\it Upper panels}: Lorentz factor in the jet along the
  axis as a function of radius, at the end of the simulation (t = 11
  s after core bounce). {\it Lower panels}: Comparison of the energy flux in the
  magnetar wind at small radii near the neutron star (solid line),
  energy flux in the jet at large radii (dotted) and energy flux in
  the cocoon at large radii
  (dashed). The jet is defined to be within $5^\circ$ of the axis,
  while the cocoon is within $20^\circ$ from the axis.}
\label{fig:ener}
\end{figure*}

As soon as the jet starts expanding into the lower density
circumstellar region it accelerates to high speeds. For computational
reasons, we cannot follow the acceleration of the jet for more than a
few stellar radii, and thus we cannot completely assess the asymptotic
acceleration of the outflow. For the range of radii that we do study,
we find that the Lorentz factor in the jet at late times is smaller
than the value of $\sigma$ at the base of the proto-neutron star wind
at the same time (the maximum achievable $\gamma$), but larger than
the $\gamma \simeq \sigma^{1/3}$ acceleration of the 1D
monopole. Specifically, in case A we find a Lorentz factor $\gamma
\sim 4$ near the axis of the jet when $\sigma \sim 10$ in the free
wind at $t \simeq 10$ sec; in case B, $\gamma \sim 10-15$ when $\sigma
\sim 60$ (see the top panel in Fig. \ref{fig:ener}).  The
magnetization of the jet itself at large radii is also not negligible:
in case A the jet has a magnetic field that is close to equipartition
($B^2/\gamma^2(\rho c^2 + 4P)\beta_r \simeq 1$), while in case B it is
somewhat above equipartition ($B^2/\gamma^2(\rho c^2 + 4P)\beta_r \simeq
5$). This largely accounts for the fact that the acceleration is not
100\% efficient.  Indeed, the late-time magnetic + kinetic energy in
the jet at large radii is comparable to the late-time value of the
magnetic energy in the magnetar wind at small radii. This is a
non-trivial result since the MWN tends to accumulate the flow so that
the highly magnetized late-time outflow could have been partially
mixed with the earlier less magnetized outflow.

Although the magnetar wind efficiently and rapidly creates a
collimated polar jet that allows relativistic material to escape the
host star, the overall interaction with the relatively spherical SN
ejecta is much weaker; in particular, little of the magnetar spin-down
power is transferred to the spherical SN ejecta. This is true even at
late times when the energy flux in the free magnetar wind at small
radii is largely equatorial.  The reason is that the wind undergoes a
termination shock and then escapes via the polar channel.  The low
equatorial pressure in the MWN leads to very little energy transfer to
the ejecta near the equator.  In case A, we do find that the MWN is
able to partially compress the shocked SN ejecta in the first few
seconds, but as soon as the jet develops the internal energy of the
MWN escapes via the polar channel and the SN ejecta start to recede
inside and partially recompress the MWN.  However, in neither case A
nor case B do we find any significant changes in the global properties
of the SN shock inside the progenitor star due to the magnetar wind.

To quantify this, Figure~\ref{fig:ener} shows, for cases A and B, the
relation between the instantaneous power in the magnetar wind at small
radii, the energy flux in the relativistic core of the jet at large
radii (defined to be the energy flux within $5^\circ$ of the pole) and
the energy flux in the wider angle wind at large radii (defined to be
the energy flux within $20^\circ$ of the pole).  Figure~\ref{fig:ener}
demonstrates that essentially all of the energy injected by the
central engine is carried away in the collimated jet with a
significant fraction of the energy confined in the central
relativistic core.  Note that there are $\sim 10-20 \%$ fluctuations
in the energy flux due to turbulence created in the curved TS near the
axis; there are significant fluctuations in the Lorentz factor as well
(top panel).  It is also interesting to note that the energy flux in
the jet and wind at large radii is on average a little higher than the
energy flux in the wind at small radii at the same time.  This is
because energy injected by the wind at earlier times can be stored in
the MWN, and released later on. The MWN thus introduces a small delay
between conditions in the wind at injection and conditions in the jet
emerging from the star.

The vorticity created at the curved TS in the polar region leads to
the development of a layer where sausage modes grow unstable. It is
reasonable to expect that in a more realistic (e.g., 3D) calculation
these instabilities might lead to the dissipation of some of the
toroidal field.  In \S \ref{sec:dis} we will discuss how these
instabilities might modify the evolution of the system.

%%%%%%%%%%%%%%%%%%%%%%%%%%%%%%%%%%%%%% FIG 4 %%%%%%%%%%%%%%%%%%%%%%%%%%%%%%
%%%%%%%%%%%%%%%%%%%%%%%%%%%%%%%%%%%%%%%%%%%%%%%%%%%%%%%%%%%%%%%%%%%%%%%%%%%%

\subsection{Case C: Lower Spindown Power}
\label{sec:casec}

\begin{figure*}
\resizebox{\hsize}{!}{\includegraphics[bb=50 70 450 520, clip]{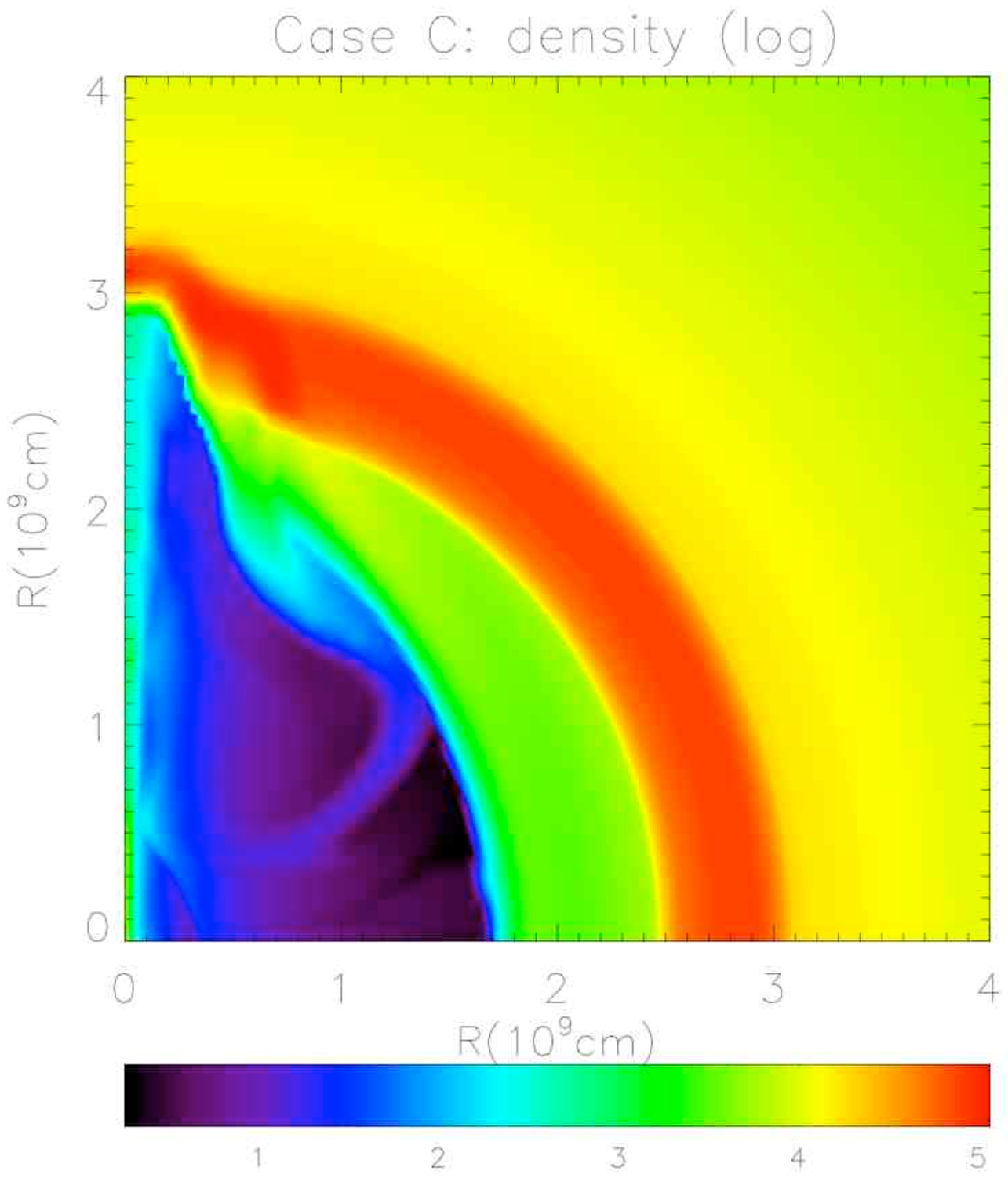}\includegraphics[bb=50 70 450 520, clip]{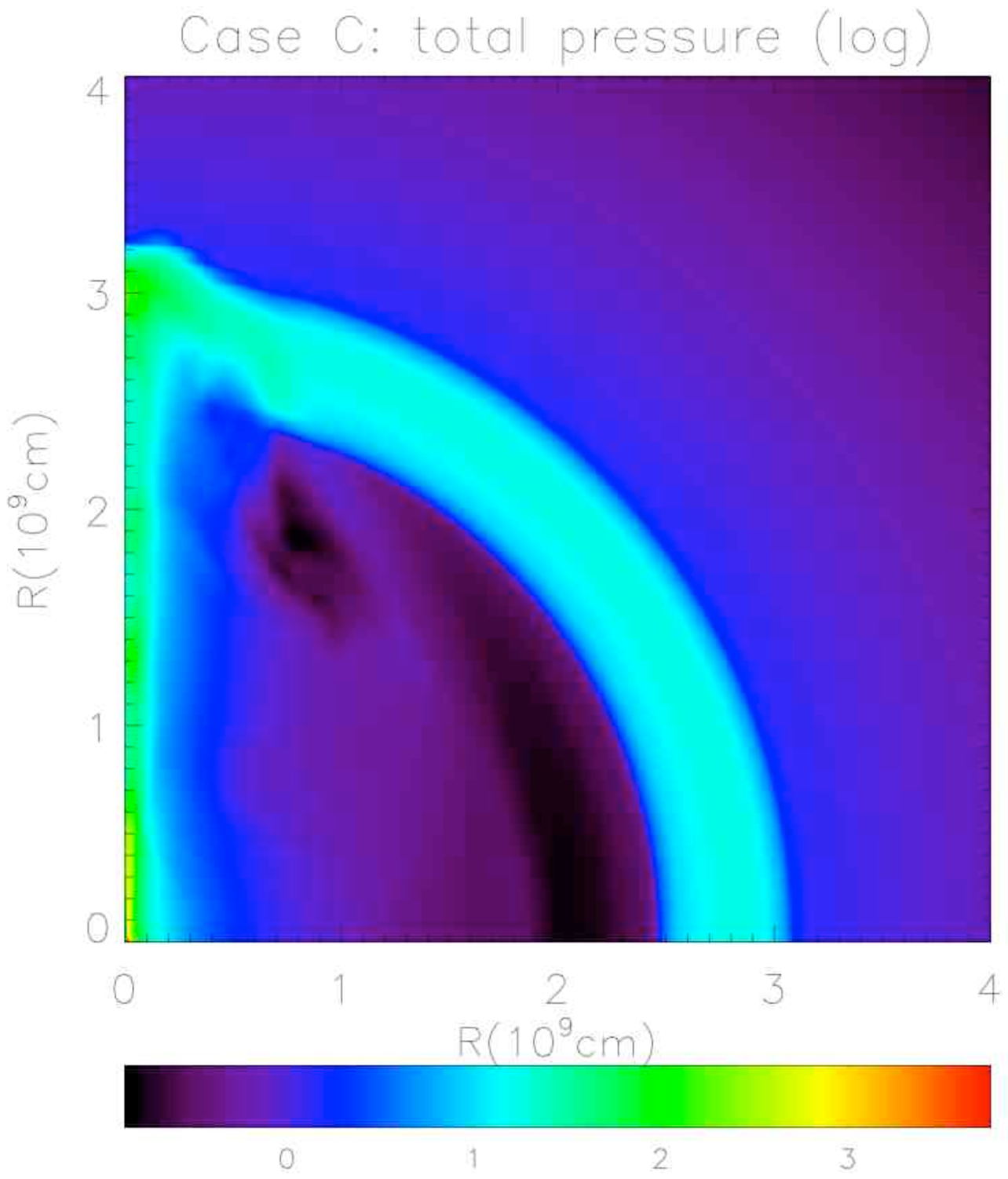}\includegraphics[bb=50 70 450 520, clip]{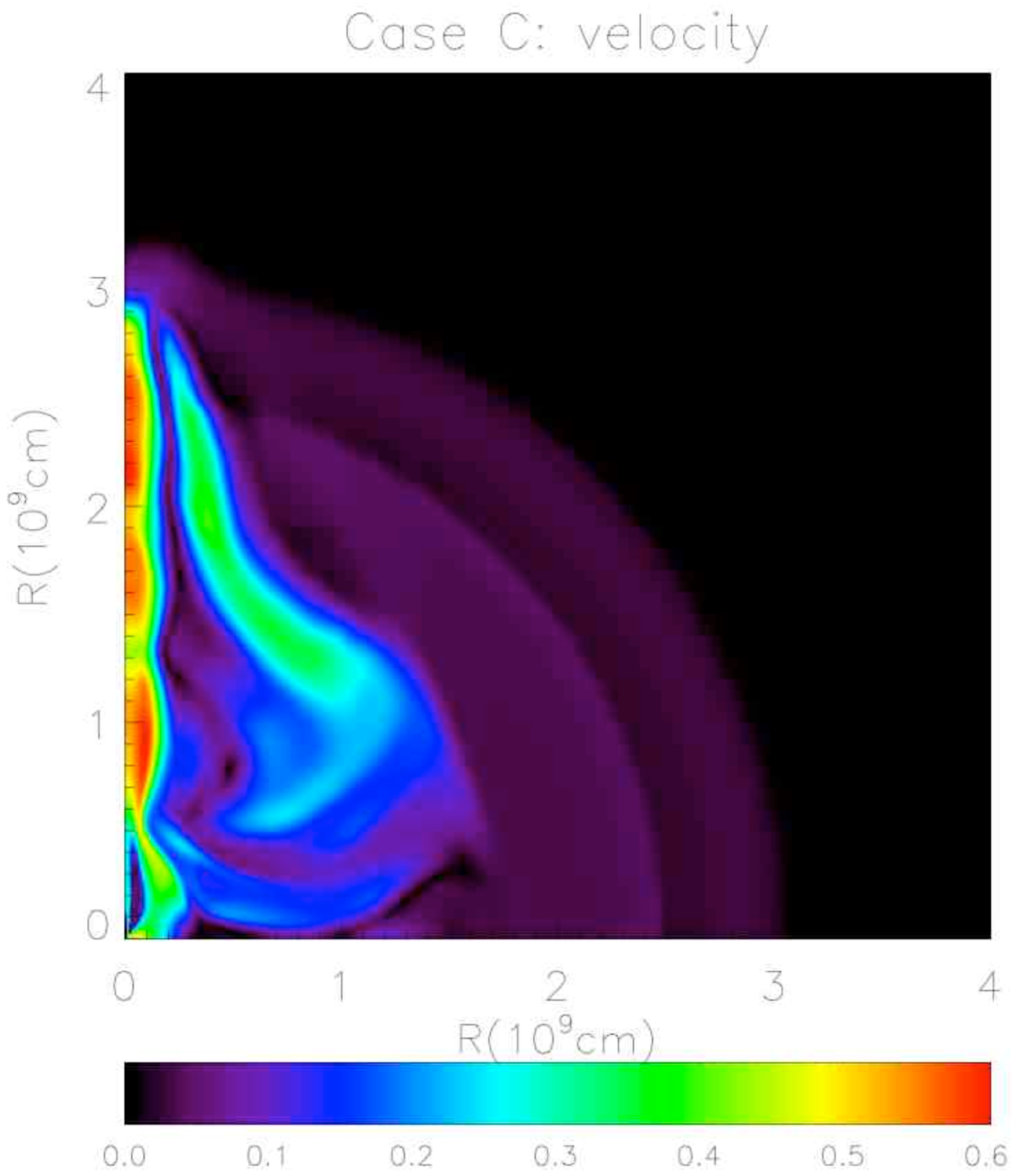}}
\resizebox{\hsize}{!}{\includegraphics[bb=50 70 450 520, clip]{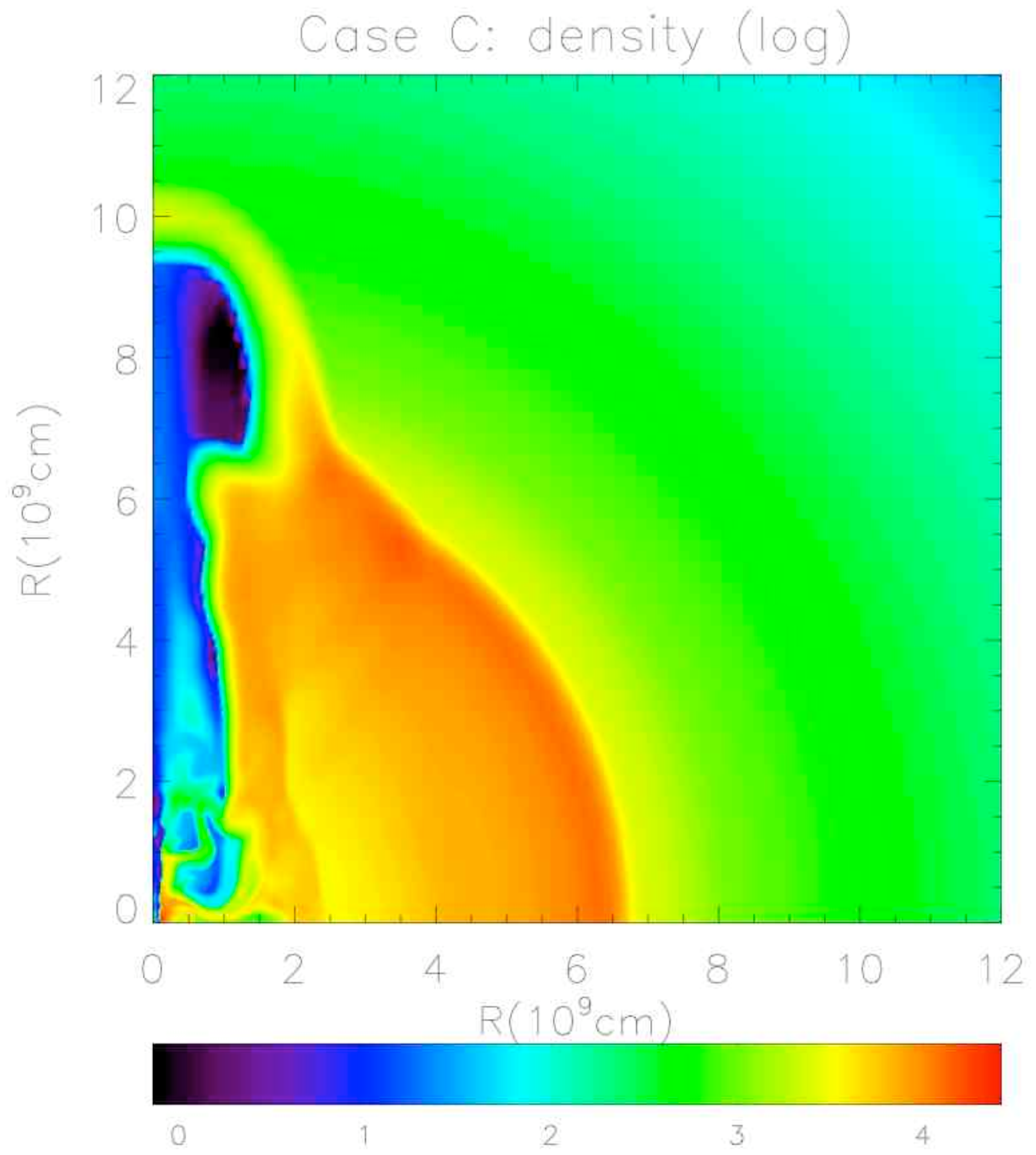}\includegraphics[bb=50 70 450 520, clip]{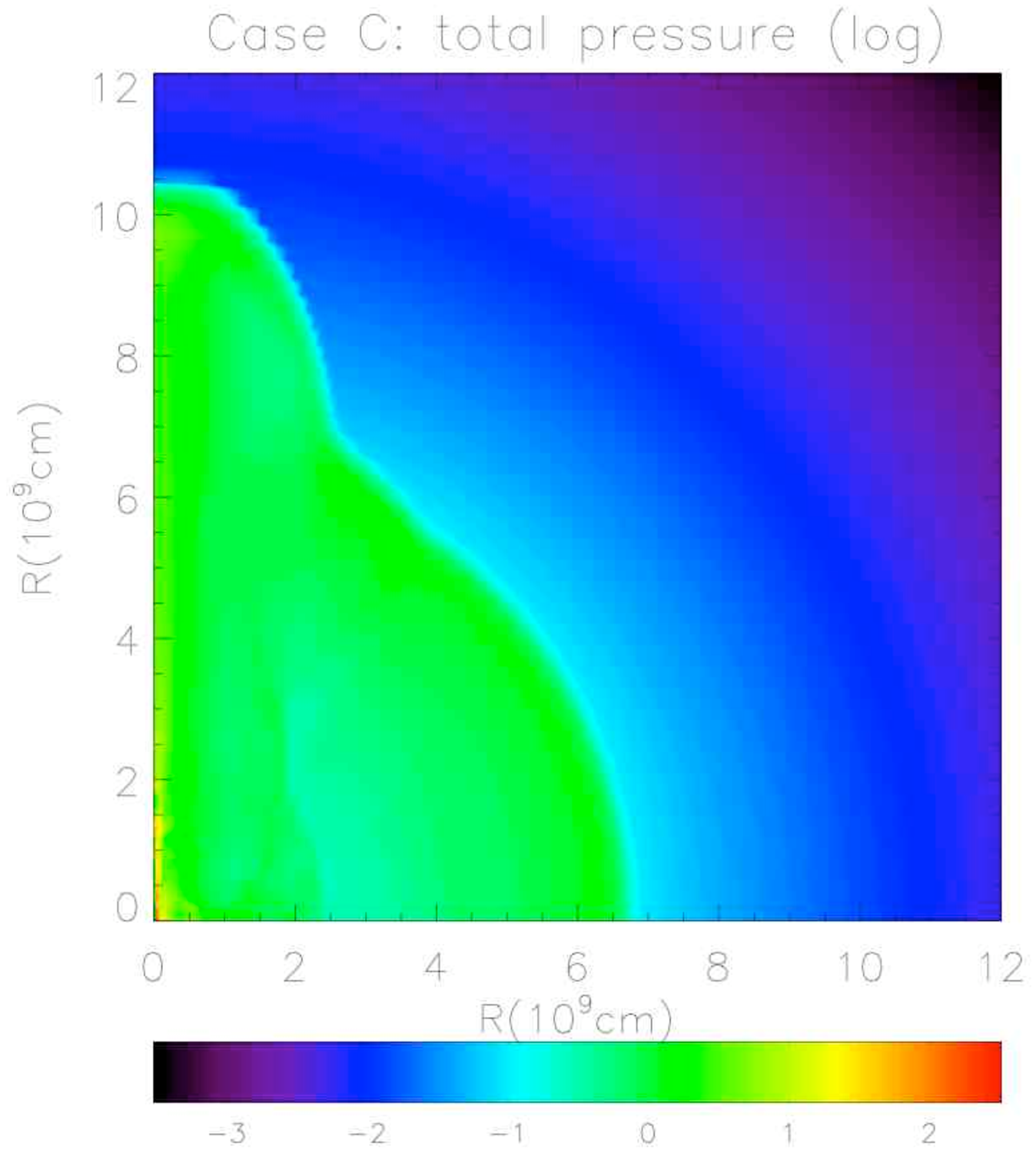}\includegraphics[bb=50 70 450 520, clip]{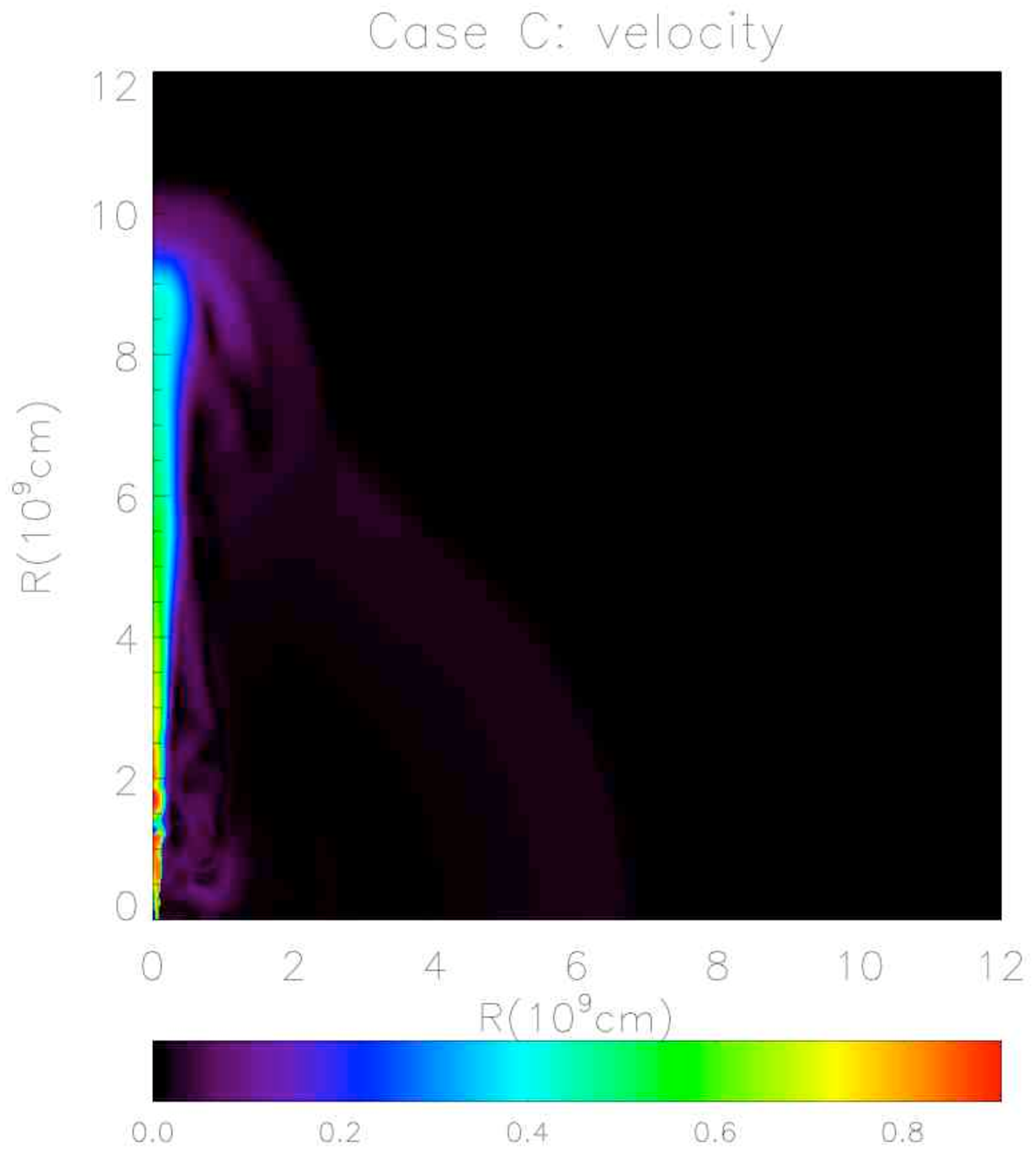}}
\resizebox{\hsize}{!}{\includegraphics[bb=50 70 450 520, clip]{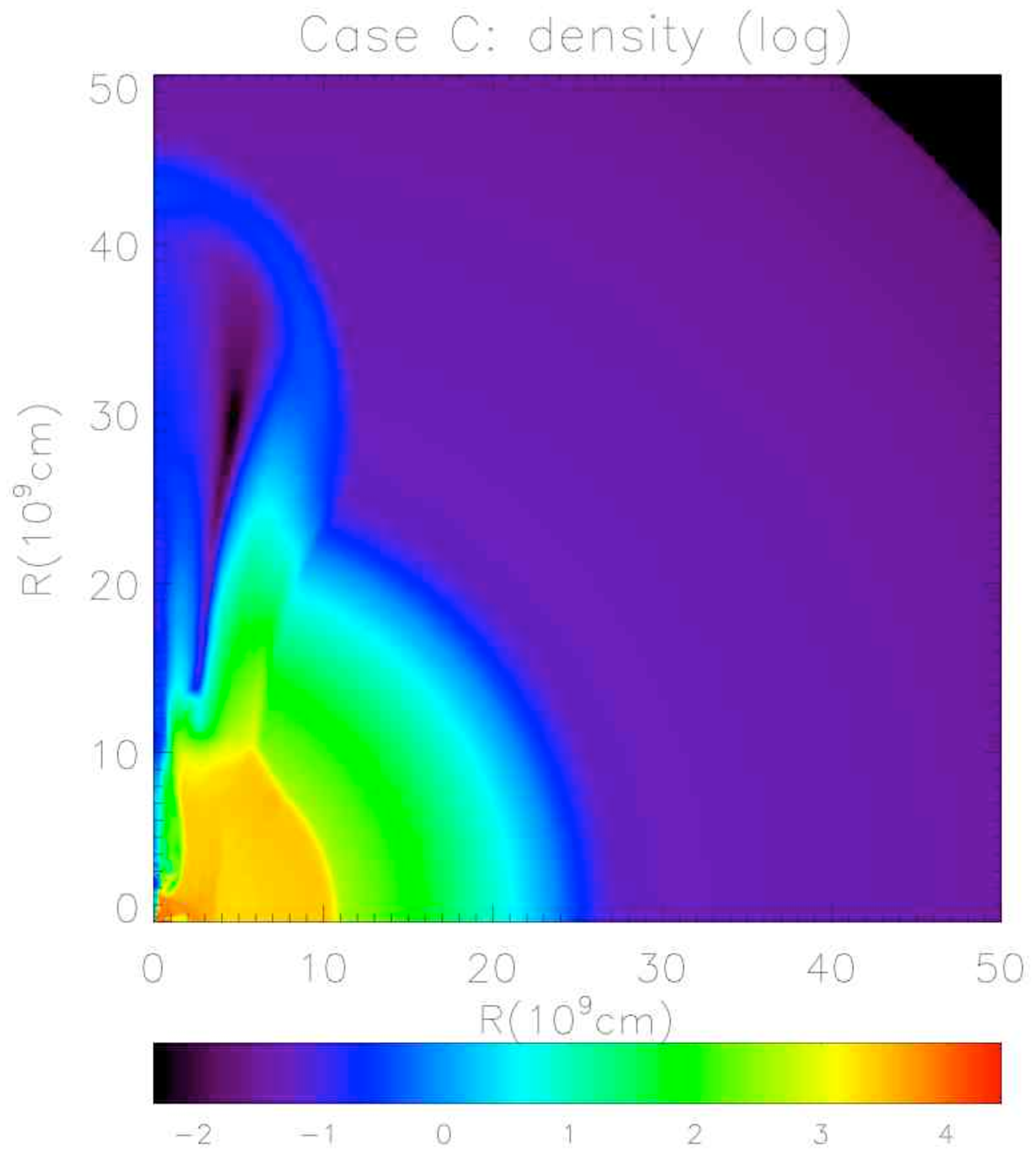}\includegraphics[bb=50 70 450 520, clip]{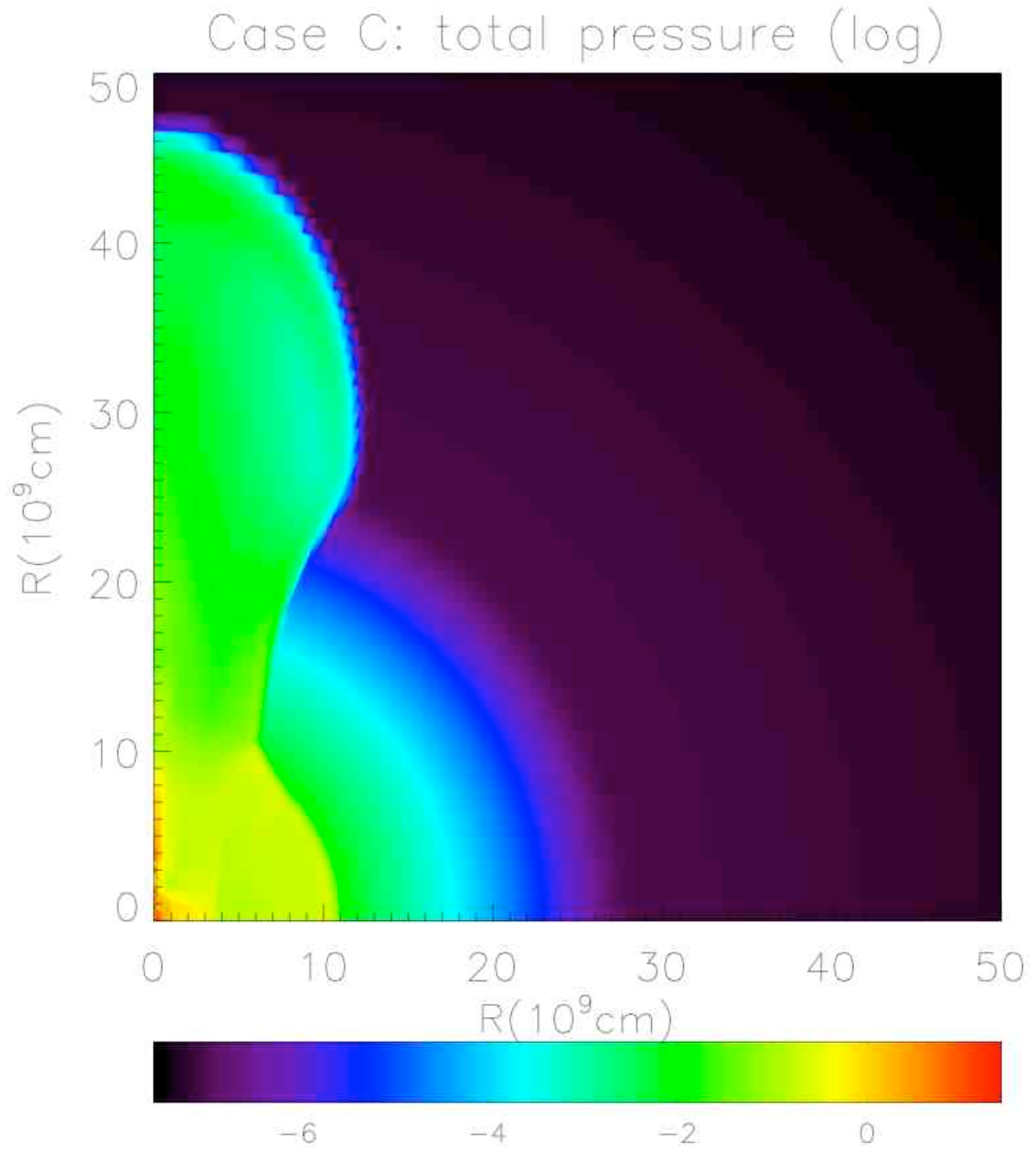}\includegraphics[bb=50 70 450 520, clip]{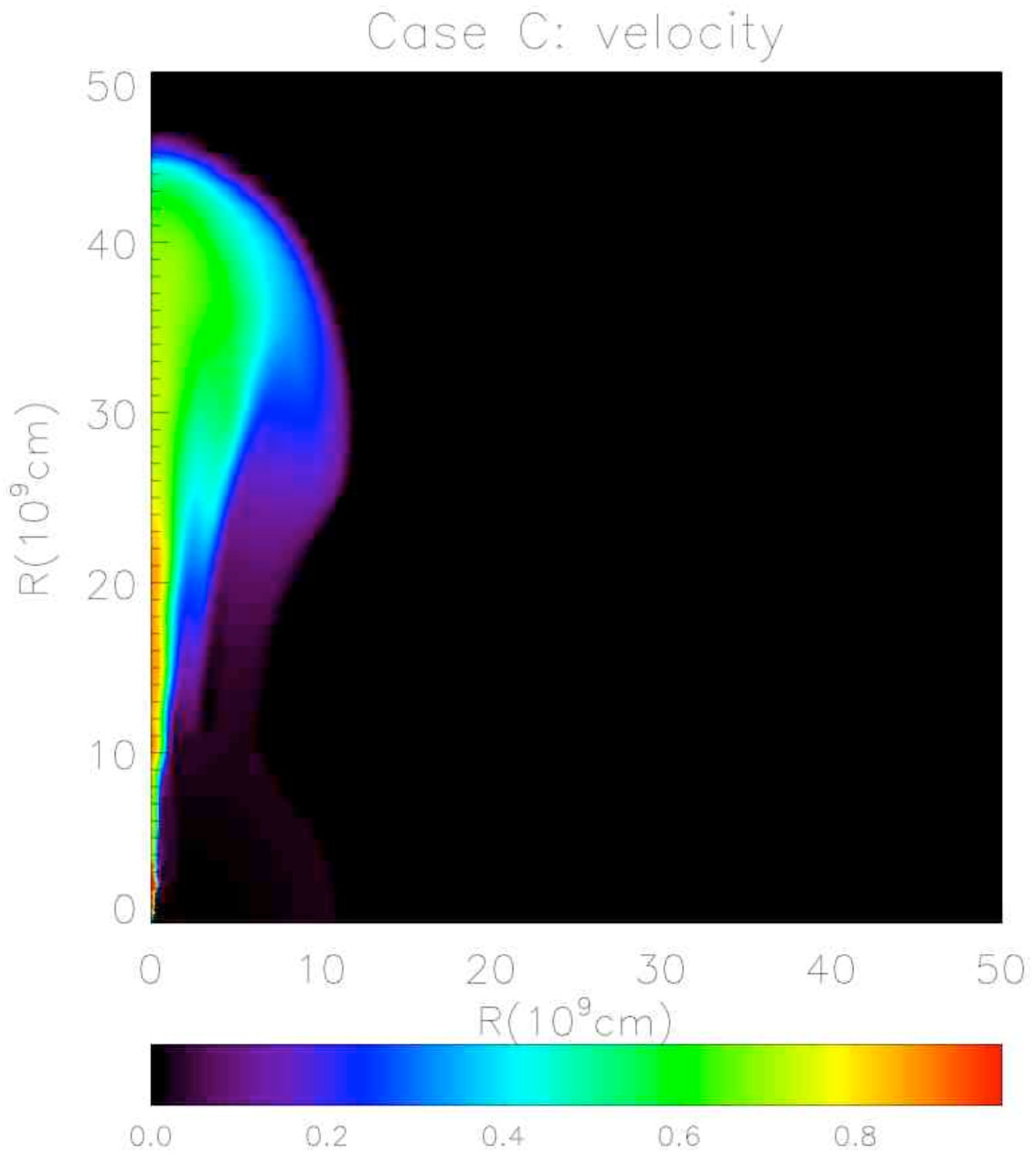}}
\caption{Evolution of the magnetized bubble inflated by a magnetar
  with $B = 10^{15} {\rm G }$ and $P = 3$ms inside a 35
  $M_\odot$ star (Case C). From left to right, the panels show:
  log$_{10}$[density (g cm$^{-3}$)], log$_{10}$[pressure (erg
  cm$^{-3}\, c^2$)] and velocity (in units of $c$). From top to
  bottom, the snapshots are $2, 6,$ \& $11$ seconds after core
  bounce. Distances are in units of $10^9$ cm; the radius of the
  progenitor star is $2.5 \, 10^{10}$ cm.  By $t \sim 10$ sec (bottom
  panel) the jet has just started to escape the progenitor star.}
\label{fig:caseC}
\end{figure*}

We now discuss case C, which corresponds to a magnetar with $P = 3$ ms
and $B = 10^{15}$ G (see Fig. \ref{fig:caseC}).  
This case differs from the previous two in two important ways: first,
the typical spindown power is a factor of $\sim 10$ ($\sim 100$) lower
than in case A (Case B); and, second, the location of the light
cylinder and thus the fast surface is at a larger distance. Because of
the lower spindown power, the evolution of the MWN proceeds more
slowly than in Cases A and B.  Ten seconds after the launching of the
wind, the MWN has barely emerged from the progenitor star.

\begin{figure*}
\resizebox{\hsize}{!}{\includegraphics[bb=50 70 450 520, clip]{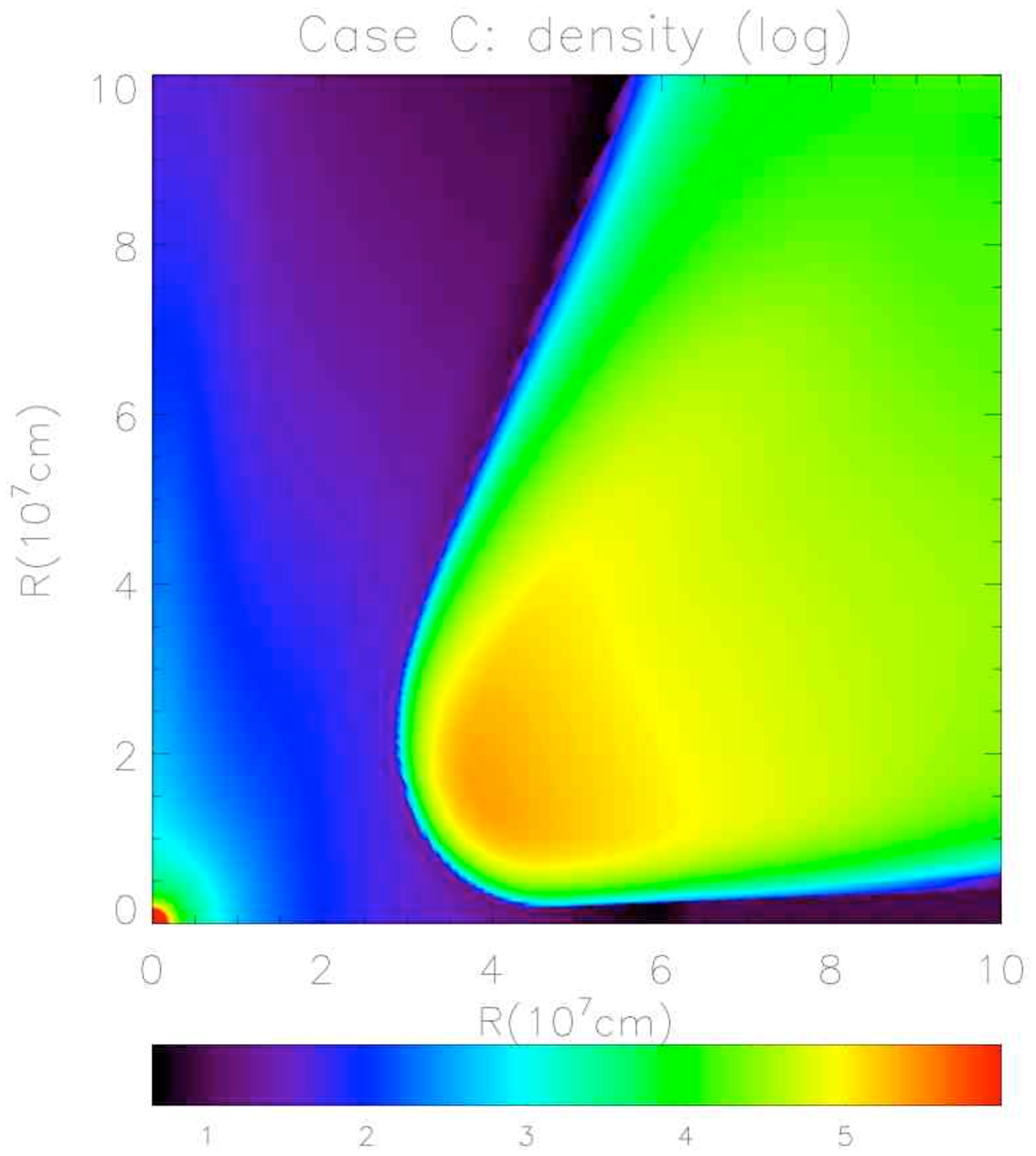}\includegraphics[bb=50 70 450 520, clip]{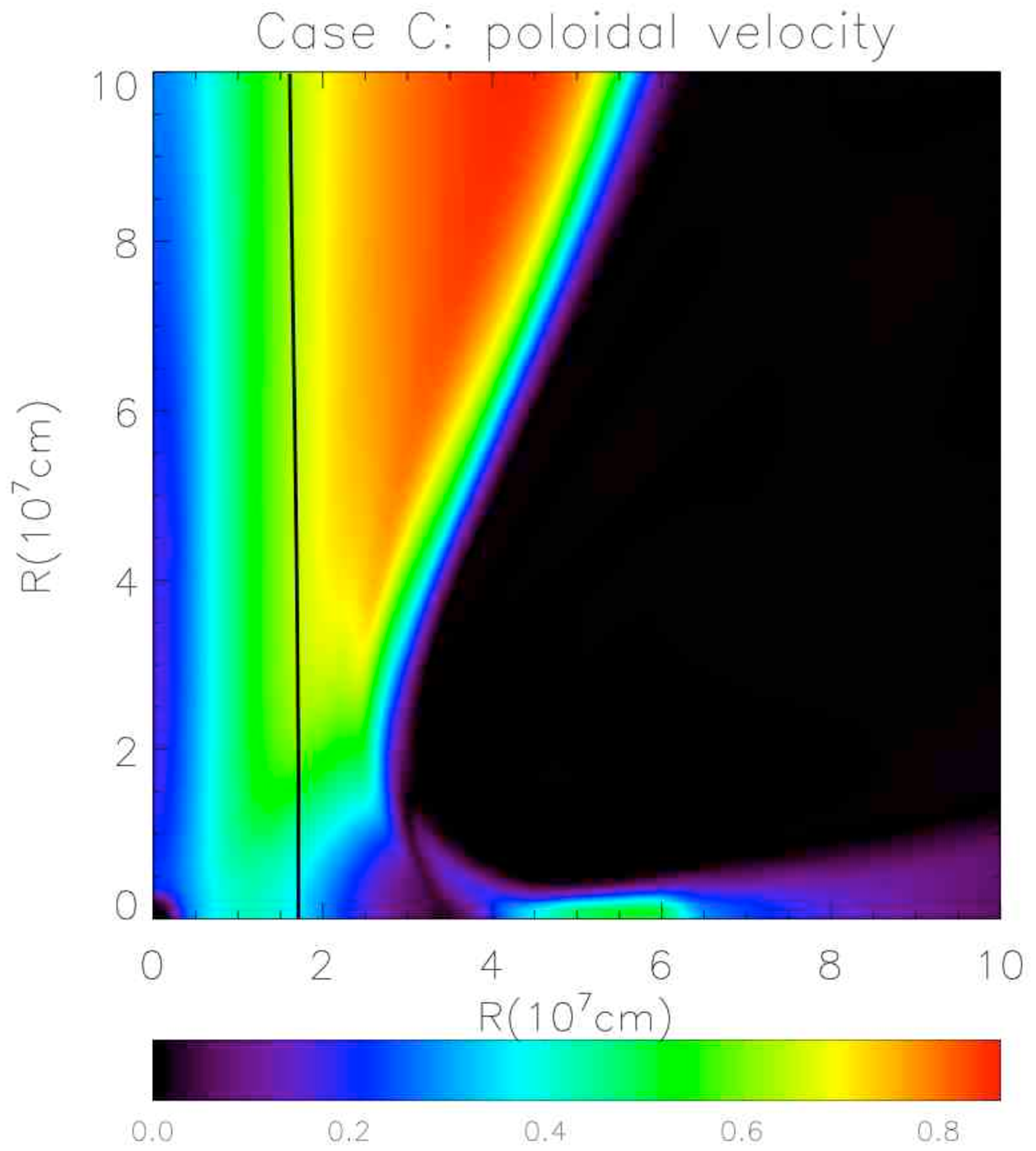}}
%\resizebox{\hsize}{!}{\includegraphics[bb=50 70 450 %520]{fig8aa_col.ps},\includegraphics[bb=50 70 450 520]{fig8ba_col.ps}}
%\resizebox{\hsize}{!}{\includegraphics{fig2b.ps}}
%\resizebox{\hsize}{!}{\includegraphics{fig3.ps}}
\caption{Density (Left panel) and poloidal velocity (in units of c;
  Right panel) in the inner $10^8$ cm near the end of our simulation
  of Case C ($B =  10^{15} {\rm G }$ and $P = 3$ms).  The
  vertical solid line near a cylindrical radius of 170 km in the
  velocity plot is the location of the Alfv\'en surface.  Note that
  the dense and slow moving stellar ejecta has compressed and
  redirected the magnetar's wind almost down to the Alfv\'en surface.
  This leads to the enhanced torque and larger spindown variability
  seen in Fig. \ref{fig:losses}}
\label{fig:zoomC}
\end{figure*}

The evolution of the MWN is also qualitatively different from the
higher spindown power cases. The lower pressure in the MWN implies
that the nebula expands at a significantly slower speed.  As a result,
even though the pressure along the rotation axis is larger than that
at the equator, the expansion of the polar jet is not fast enough to
overcome the compression of the toroidal field in the nebula, and so
the termination shock collapses to smaller radii (unlike in the higher
spindown cases where the rapid polar expansion allows the system to
avoid this fate).
A sub-fast flow is first established in the polar region, but after
about 1 second the termination shock contracts within the fast surface
along the equator as well, resulting in a fully sub-fast outflow. The
dynamics in this case is qualitatively similar to that found by
\citet{kom07} although the absolute radial scales are larger. In \S
\ref{sec:torque}, we will discuss the implications of the sub-fast
outflow for the spindown of the proto-magnetar.  As in the previous
cases, vorticity is created downstream of the termination shock that
generates turbulence which persists to late times.  This turbulence
gives rise to significant interpenetration of the wind and the SN
ejecta, in the form of fingers of SN ejecta that are dragged down to
small radii, as is shown in the zoom-in to small radii in Figure
\ref{fig:zoomC}. At late times, the denser SN ejecta even compress the
magnetar wind at radii close to the Alfv\'enic surface.

%%%%%%%%%%%%%%%%%%%%%%%%%%%%%%%%%%%%%% FIG 5 %%%%%%%%%%%%%%%%%%%%%%%%%%%%%%
%%%%%%%%%%%%%%%%%%%%%%%%%%%%%%%%%%%%%%%%%%%%%%%%%%%%%%%%%%%%%%%%%%%%%%%%%%%%

We have attributed the differences between cases A/B and case C --
namely a largely super-fast vs. fully sub-fast outflow -- to the
differences in spindown power and rotation rate.  One concern might be
whether these properties of the outflow are sensitive to the size of
the ``SN'' cavity we initialize. To assess this, we have repeated case
A using a cavity with an initial radius of $5 \, 10^8$ cm, a factor of
2 smaller than in our fiducial case shown in Figure \ref{fig:caseA}.
At very early times $\simlt 1$ sec the smaller cavity does modify the
solution, leading to a fully sub-fast outflow because the termination
shock is at smaller radii.  However, as the system evolves and the MWN
expands into the outgoing SN ejecta a super-fast outflow is again
established in the equatorial region.  After $\sim 1$ sec, the system
relaxes to a configuration similar to that which we found starting
with a larger initial cavity.  This demonstrates that the primary
physics determining whether the magnetar wind is super-fast or
sub-fast is indeed the spindown power of the neutron star.

%%%%%%%%%%%%%%%%%%%%%%%%%%%%%%%%%%%%%% FIG 6 %%%%%%%%%%%%%%%%%%%%%%%%%%%%%%
%%%%%%%%%%%%%%%%%%%%%%%%%%%%%%%%%%%%%%%%%%%%%%%%%%%%%%%%%%%%%%%%%%%%%%%%%%%%

\subsection{Proto-Neutron Star Spindown Rate}

\label{sec:torque}

In all of our simulations, there is at least some fraction of the
proto-neutron star outflow that is in causal contact with the
surrounding stellar envelope.  In the higher spindown power
simulations (cases A and B), the polar region is sub-fast, while in
the lower spindown power simulation (case C), the entire outflow is
sub-fast.  In all three cases the outflow is always super-Alfv\'enic.
Because of this causal contact, it is possible that the spindown of
the neutron star could be modified from that due to a free wind.
Understanding the torque on the central engine is important because it
strongly influences whether or not conditions suitable for producing
GRBs can be achieved. For example, an increased torque due to a
sub-fast outflow might lead to sufficiently rapid spindown that most
of the rotational energy has been depleted before the outflow becomes
relativistic.

%%%%%%%%%%%%%%%%%%%%%%%%%%%%%%%%%%%%%% FIG 5 %%%%%%%%%%%%%%%%%%%%%%%%%%%%%%
\begin{figure*}
\resizebox{\hsize}{!}{\includegraphics[bb= 90 395 570 1165, clip]{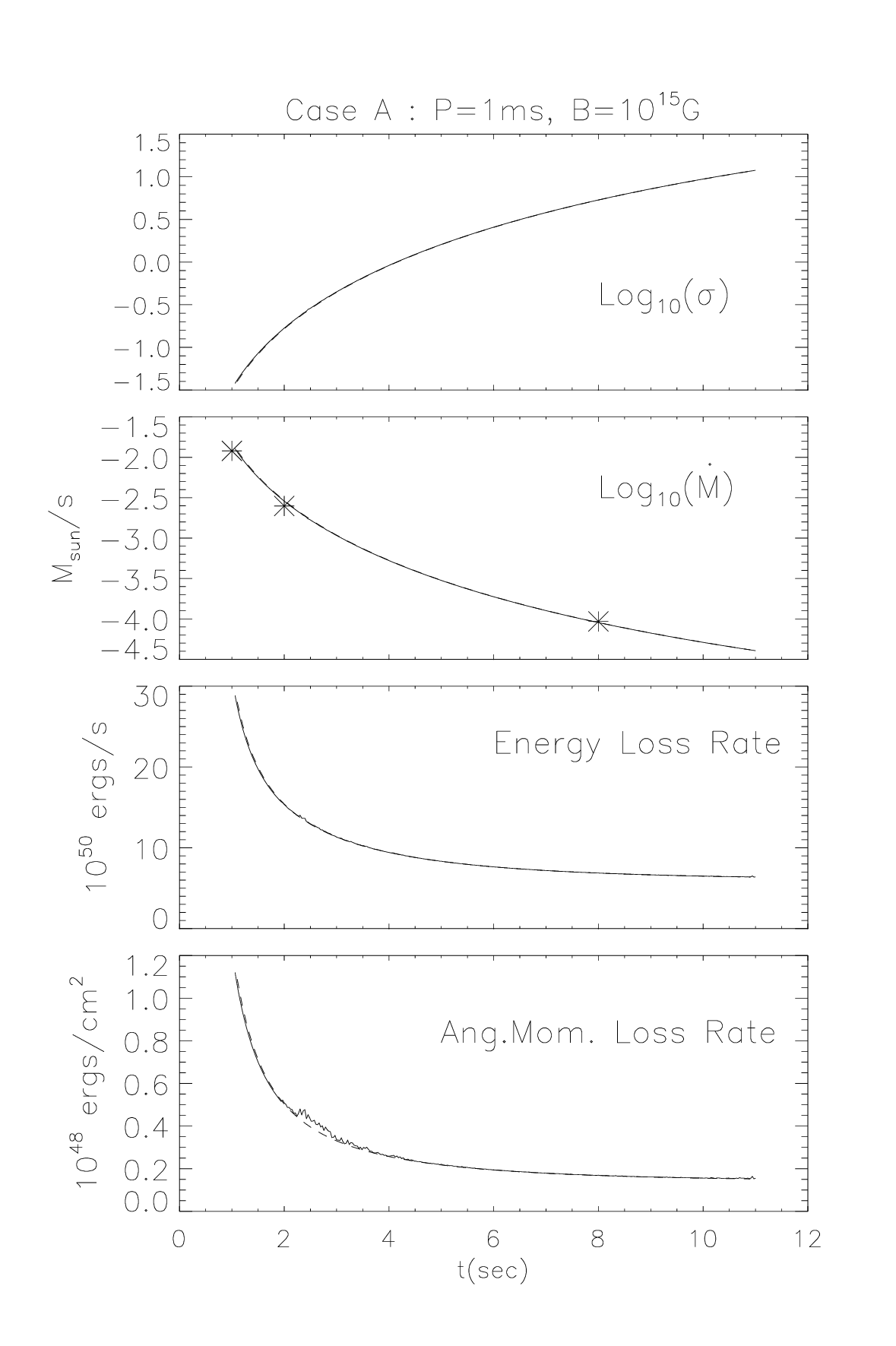}\includegraphics[bb= 90 395 570 1165, clip]{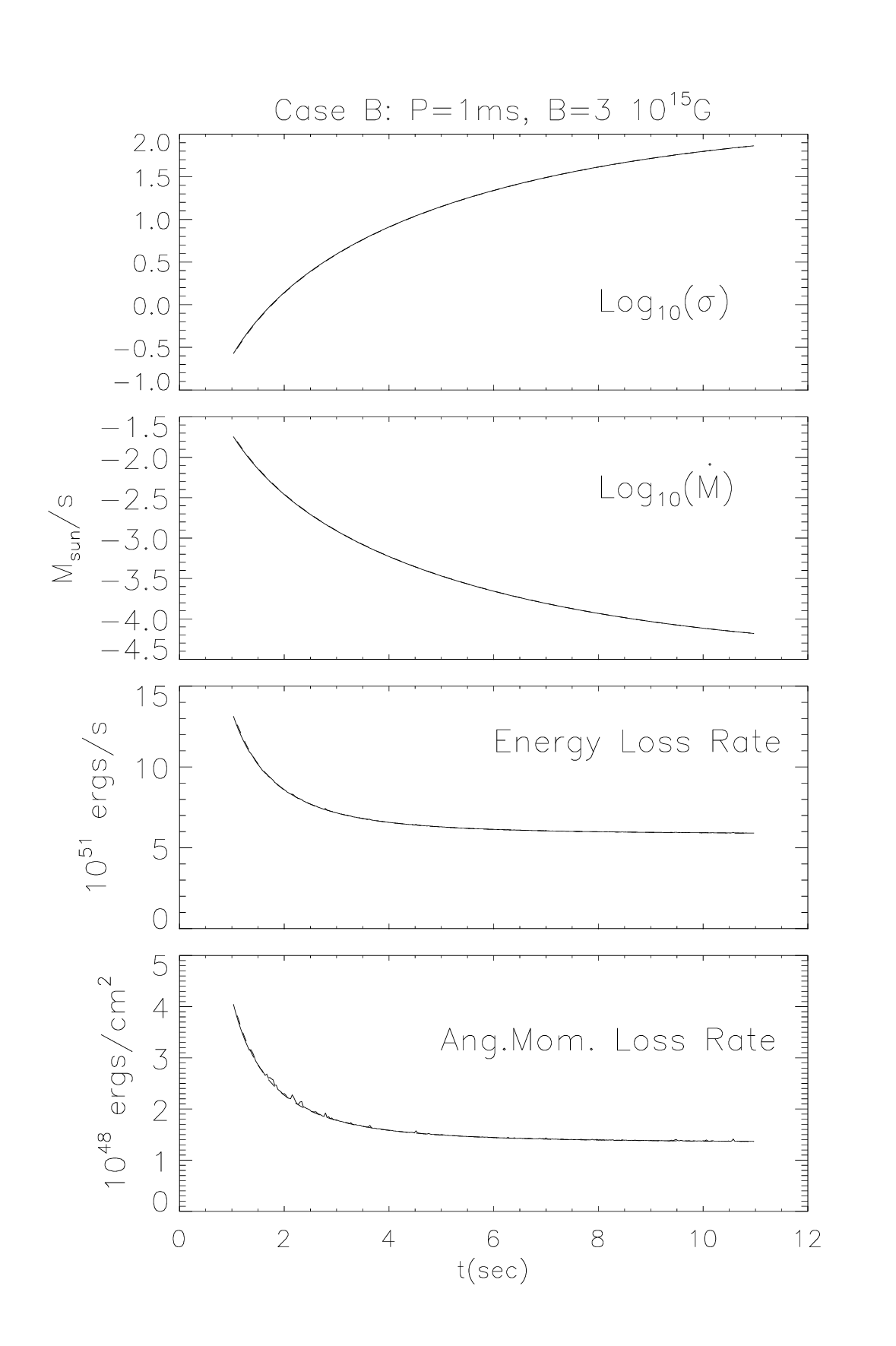}\includegraphics[bb=90 395 570 1165, clip]{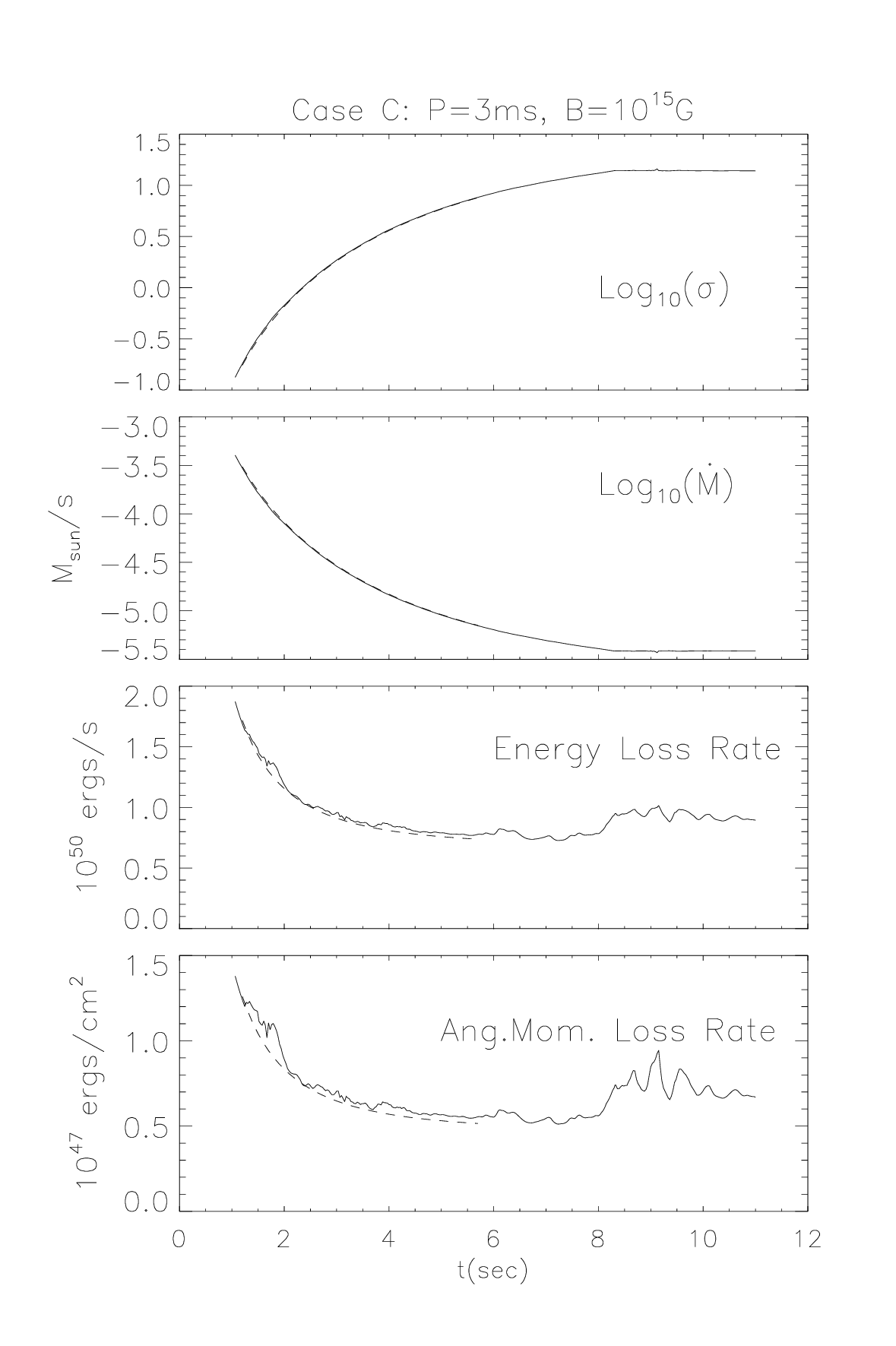}}
\caption{Magnetization $\sigma$ and mass, angular momentum, and energy
  losses for cases A (left), B (center), and C (right). The dashed
  line represents the same quantities computed for the free wind; for
  case C, the free wind was numerically unstable at late times and so
  is not plotted after $\simeq 6$ s (\S \ref{sec:num}).  For case A,
  we also show the mass loss rate at 3 times (stars) from the
  calculations of \citet{met07}, who included realistic neutrino
  microphysics.}
\label{fig:losses}
\end{figure*}
%%%%%%%%%%%%%%%%%%%%%%%%%%%%%%%%%%%%%%%%%%%%%%%%%%%%%%%%%%%%%%%%%%%%%%%%%%%%

Figure \ref{fig:losses} shows the magnetization, mass loss rate,
energy loss rate and angular momentum loss rate for the central
neutron star in our simulations (solid) compared to the corresponding
results for free winds (dashed lines). Figure \ref{fig:losses} shows
that for cases A and B, the energy and angular momentum losses in the
wind are essentially identical to those of a free wind. At first, this
seems to contradict the fact that a sub-fast wind has a larger torque
than a super-fast wind in 1D calculations.  However, for cases A and B
the sub-fast outflow is present only in the polar region, while at
high $\sigma$ most of the torque is exerted by the equatorial
outflow. The polar outflow thus makes only a small contribution to the
total torque even if it causally connected with the envelope.

The close similarity between the torque in our simulations and that
exerted by a free wind can also be understood in terms of the location
of the critical surfaces.  In 1D there are 2 free parameters at the
inner boundary (i.e. $B_\phi$ and $v_r$), and 2 corresponding critical
points, the slow and fast magnetosonic points; the equations can be
renormalized to eliminate the singularity at the Alfv\'en point, which
is automatically crossed once the solution passes smoothly through the
slow and fast points.  These two critical points correspond to two
eigenvalues of the system: the mass loss rate, associated with the
slow point, and the terminal Lorenz factor, associated with the fast
point (the torque, which is also an eigenvalue, is determined by the
Alfv\'en point, and, in 1D, is fixed once the other two are known).
In 2D, however, there are three free parameters at the inner boundary
($B_\theta$, in addition to $v_r$ and $B_\phi$) and 3 corresponding
critical surfaces and 3 independent eigenvalues: the slow, fast, and
Alfv\'en surfaces.  In order to modify the torque significantly, the
location and shape of the Alfv\'en surface must be modified.  For fast
rotators like those considered here, this implies that the outflow
must be modified inside the light cylinder.  We find, however, that
even if the termination shock moves inside the fast surface, the
location of the Alfv\'en surface remains relatively unchanged.

The results for case C -- in which the outflow is fully sub-fast --
are somewhat different than those for Cases A/B.  Although the torque
and energy losses are nearly identical to that of the free wind at
early times, at late times the energy losses tend to be more variable
and can be $\sim 30\%$ higher than for the free wind
(Fig. \ref{fig:losses}).  These large fluctuations correspond to when
the post termination shock turbulence is able to compress the
proto-magnetar wind down to radii that are close to the light cylinder
(Fig. \ref{fig:zoomC}).  This level of fluctuations, while interesting
from the point of view of the underlying physics of the problem, is
unlikely to change the overall evolution of the system in a
significant way.

\subsection{Post-Shock Nucleosynthesis}

\label{sec:nuc}
  
%%%%%%%%%%%%%%%%%%%%%%%%%%%%%%%%%%%%%% FIG 7 %%%%%%%%%%%%%%%%%%%%%%%%%%%%%%
\begin{figure}
\resizebox{\hsize}{!}{\includegraphics[bb=50 70 450 520, clip]{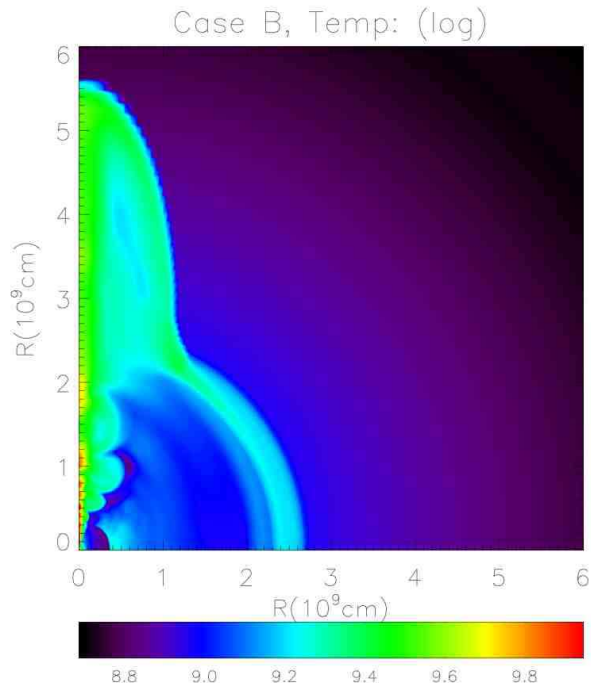}}
\caption{Temperature (log$_{10}$[T (K)]) for case B at t=1.6 sec after
  core bounce.  The temperature is never high enough for significant
  nucleosynthesis of $^{56}$Ni, which requires $T \simgt 5 \, 10^9$ K,
  but some nucleosynthesis of Carbon and Oxygen will occur in material
  shock heated by the polar jet.}
\label{fig:temp}
\end{figure}
%%%%%%%%%%%%%%%%%%%%%%%%%%%%%%%%%%%%%%%%%%%%%%%%%%%%%%%%%%%%%%%%%%%%%%%%%%%%

Our calculations have assumed an ideal gas equation of state with an
adiabatic index of $4/3$, which is appropriate for a radiation
dominated flow or for relativistic conditions.  We have checked a
posteriori that this assumed adiabatic index is consistent with the
thermal properties of the MWN and jet.  Figure~\ref{fig:temp} shows
the temperature derived from inverting the relation: \be p={k_B \rho T
  \over \mu m_p} + {a T^4 \over 3} \ee between the pressure and
density in our simulations, at t = 1.6 sec after core bounce.  At
almost all locations, the temperatures are high enough that radiation
pressure dominates and thus our equation of state is self-consistent.
 
The temperature of the gas is also important for understanding
nucleosynthesis during the SN explosion.  For example, explosive
nucleosynthesis of $^{56}$Ni requires temperatures of $\simgt 5 \,
10^9$ K \citep{woo02}.  Figure \ref{fig:temp} shows that this
temperature is not attained even at relatively early times in our
simulations.  The reason for this is the low density of the outer
stellar envelope. By the time the jet-plume emerges outside the SN
shock the density of the progenitor is $\sim 10^{4-5}$ g cm$^{-3}$. At
these densities Ni production requires a shock moving at nearly the
speed of light, significantly faster than what we find even for our
most energetic explosion (case B).  We do find, however, that $\sim
10^{-2} M_\odot$ of high speed ($v \simeq 0.1-0.2 \, c$) Ne and O can
be created, because these have lower threshold temperatures for
successful explosive nucleosynthesis.  The synthesized mass is
relatively low because only a small solid angle near the pole is
shock-heated to sufficiently high temperatures by the jet. In the
context of observed GRBs, our viewing angle is relatively {\it
  on-axis}, in which case the high velocity nucleosynthesis may be
observable; high velocity O and Ne can also be produced by the jet
blowing out stellar material that had been processed during stellar
evolution \citep{maz06}. Even if the jet nucleosynthesis only
contributes marginally to the total nucleosynthesis during the
explosion, it might lead to unique observable signatures in the ejecta
at late times (as may be the case for Cas A; \citealt{whe08}). 

It is clear that the jet could interact with higher density gas, and
thus be more likely to produce Ni, at early times $\simlt 1$ sec, when
the SN shock and MWN are smaller.  However, the physics at these early
times is uncertain: the magnetar is likely still contracting to its
final radius and the magnetic field may even still be growing via a
dynamo.  Moreover, the early time evolution is likely to be more
sensitive to the details of the explosion mechanism.  As described
previously (\S \ref{sec:torque}), we have verified that the late-time
spindown of the proto-neutron star and the formation of the collimated
jet, are not sensitive to the size of the initial cavity we choose
(which is a proxy for the uncertain SN physics).  This will not be
true of the amount of Ni synthesized.  We thus conclude that a more
careful treatment of the early time contraction and spindown of the
magnetar, probably together with a better understanding of the SN
mechanism, are required to reliably assess the question of whether the
central engine responsible for the GRB also contributes significantly
to the production of Ni during long-duration bursts.

\section{Discussion \& Implications}
\label{sec:dis}

The calculations presented in this paper, together with our previous
work (Papers I \& II), show that the production of a collimated
relativistic jet that can escape the progenitor star is a robust
consequence of the formation of a $B \sim 10^{15}$ G, $P \sim 1$ ms
neutron star during core-collapse supernovae.  Paper II shows that
such a jet is formed in the low $\sigma$ limit in which most of the
magnetic energy in the magnetar's outflow has been converted into bulk
kinetic energy -- a limit motivated by observations of pulsar wind nebulae which
show that such efficient conversion takes place (e.g.,
\citealt{ken84}).  In this paper we have focused on the opposite
limit, that of a highly magnetized outflow in which the only
conversion of magnetic energy into kinetic energy is that which takes
place in ideal MHD.  Just as in the low $\sigma$ limit, we find that a
bipolar jet begins to form in the first $\sim 1$ sec after
core-collapse and escapes the star after $\sim 3-10$ sec (depending on
the exact values of $B$ and $P$; see Figs
\ref{fig:caseA}-\ref{fig:caseC}).  Given the uncertainties in the
conversion of magnetic energy into kinetic energy in magnetized
outflows, it is reassuring that the properties of the GRB-producing
jet are {\it not} that sensitive to these details.  Physically, this
is because in all cases the toroidal magnetic field produced by the
central neutron star builds up in the cavity created by the outgoing
SN shock, until the field is sufficiently strong to drive the flow out
in the polar direction; because the speed of the SN shock $\sim
0.03-0.1$ c is much less than the speed of the magnetar wind $\sim$ c,
such a build up of toroidal field is inevitable unless the magnetic
energy in the wind is extremely small ($\sigma \ll 0.01$).

An important result of our work is that nearly all of the spindown
energy of the neutron star escapes via the polar channel
(Fig. \ref{fig:losses}).  There is very little energy transferred to
the exploding star in the equatorial region.  We again find this in
both the low (Paper II) and high (this paper) $\sigma$ limits. One
implication of this result is that the central engine powering the GRB
is unlikely to contribute significantly to energizing the SN shock as
a whole (although it clearly does so in the polar region), at least on
timescales $\simgt 1$ sec after core bounce; at earlier times, the
dynamics will be sensitive to the details of the SN mechanism.  We
believe that this conclusion is not specific to the particular central
engine considered here, but will hold for all {\it
  magnetically-driven} GRB central engines.  Specifically, we suspect
that wide-angle magnetized winds from accretion disks, such as are
seen in simulations of accretion disks relevant to GRBs (e.g.,
\citealt{proga03}), will form a magnetized bubble that will escape via
the polar channel rather than transferring energy to the SN shock as
has been previously hypothesized \citep{kohri05} (see also \citep{aro03}).

Observationally, there is a strong association of long-duration GRBs
with core-collapse SN, in particular energetic SN Ic-BL (BL = broad
line) \citep{woo06,del06}.  However, the converse is not true
\citep{soderberg06,woo06}.  Late-time radio observations of large
samples of Type Ibc SN find that even the broad line subset do not
show evidence for an energetic relativistic outflow, which would
produce radio emission at late times regardless of whether or not any
putative gamma-ray emission was initially beamed towards us
\citep{soderberg06}.  More concretely, \cite{soderberg06} rule out
with 84\% confidence the hypothesis that every broad-lined SN harbors
a GRB.  Moreover, the SNe associated with long-duration GRBs are not
particularly unusual among the class of BL SN in terms of their
energies, photospheric velocities, and Ni masses (e.g.,
\citealt{soderberg06b,woo06}).  A plausible interpretation of these
data is that some not yet fully understood physics (rotation?)  leads
to a class of core-collapse SNe that are unusually energetic and
asymmetric (as revealed by spectra-polarimetry), and that produce
significant amounts of Ni.  A subset of those SNe in turn produce
relativistic jets that power GRBs, but the central engine that powers
the GRB does not significantly modify the global properties of the
coincident SNe. Clearly, the detection of a GRB implies a preferential
viewing geometry, along the rotation axis of the system; this is
likely accompanied by the detection of a modest amount of high
velocity ejecta and nucleosynthetic products (\S \ref{sec:nuc}).
However, there is no indication that the global properties or the
off-axis appearance of the SNe associated with GRBs are strongly
affected by the central engine that produces the relativistic jet.
Indeed, our results demonstrate that after $\simeq 1$ sec, most of the
energy produced by a spinning-down magnetar escapes along the polar
jet rather than energizing the SN shock.  As argued above, we suspect
that this result is likely to hold for all magnetically-driven GRB
models, including disks accreating onto black holes, but this remains
to be explicitly demonstrated.

In our calculations we have assumed that ideal MHD holds in the
magnetized bubble created behind the SN shock. It is, however, well
known that plasmas with a primarily toroidal magnetic field can be
unstable to non-axisymmetric kink modes \citep{begelman98} that would
not be captured by our axisymmetric simulations (although such
instabilities may be stabilized by rotation; see \citealt{tomi01}).
This raises obvious questions about the overall stability of the
magnetic configuration found here (see also the recent results by 
\citet{bla08} about jet stability).
Unfortunately, it is not presently feasible to carry out 3D
simulations of our magnetar wind model with the necessary spatial
resolution and temporal duration.  As noted above, however, the
production of a relativistic jet by the interaction of a magnetar wind
with the surrounding stellar envelope appears to be relatively
insensitive to the precise magnetization of the outflow -- as a
comparison of the results of this paper and those of Paper II
demonstrates.  The most significant difference is in the acceleration
of the jet once it emerges from the progenitor star: in the highly
dissipative case (Paper II) the terminal Lorentz factor is larger
($\sim \sigma$ at the Light Cylinder), while in ideal MHD (this
paper), the terminal Lorentz factor is somewhat lower, but the outflow
remains reasonably strongly magnetized ($\sigma \simgt 1$) at the
radii we can study.  We thus conclude that unresolved instabilities
are unlikely to change the large-scale evolution of the magnetar wind
bubble or the resulting production of a collimated jet, but they may
well be important in determining the acceleration of the jet and the
exact ratio of magnetic to kinetic energy in the jet at large radii.

Our results also have implications for the nucleosynthesis of heavy
elements due to the interaction between the magnetar-driven jet and
the stellar envelope; we largely confirm previous studies on
nucleosynthesis in jets and asymmetric supernovae, which were based on
parameterized models for the origin of the jet/asymmetry
\citep{mae02,nag03,nag06}.  As noted above, most of the energy
produced by the magnetar is diverted into a jet with an opening angle
of $\sim 5-10^\circ$.  The amount of mass swept-up by the jet is $\sim
0.1 M_\odot$, most of which is Oxygen in the stellar model we use
\citep{woo02}.  Even in our most energetic simulations, the
temperature at the head of the jet is insufficient to produce
$^{56}$Ni (\S \ref{sec:nuc} and Fig. \ref{fig:temp}).  It is, however,
sufficient to produce $\sim 10^{-2} M_\odot$ of high velocity Ne and
Mg, and perhaps a similar amount of high velocity Oxygen.
Unfortunately, the high velocity Oxygen that is observed in some SNe
associated with GRBs \citep{maz06} can easily be explained as part of
the stellar envelope that is blown out by the jet, and thus these
observations do not strongly constrain the properties of the material
shock-heated by the head of the jet as it propagates through the star.
The reason that our calculations do not produce a significant amount
of $^{56}$Ni is that we initialize our simulations roughly 1 second
after core bounce, at which time the SN shock has moved to $\sim 10^9$
cm and the post-shock temperature at the head of the jet is $\simlt 5
\times 10^9$ K.  The temperature could be higher at earlier times, but
the physics at earlier times, and the amount of $^{56}$Ni produced,
will likely depend sensitively on the physics of the SN explosion
itself.

Our simulations span a factor of $\sim 100$ in the neutron star's
spindown power (Fig. \ref{fig:losses}).  In our high spindown power
simulations, corresponding to neutron stars with $P = 1$ ms and $B
\sim 1-3 \times 10^{15}$ G, we find that the outflow remains largely
out of causal contact with the surrounding stellar envelope.  In
particular, unlike in 1D models of highly magnetized outflows
\citep{ken84}, in our 2D calculations we find that even for high
$\sigma$, the termination shock between the wind and the magnetar wind
bubble lies outside the wind's fast magnetosonic surface, except very
near the pole (Fig. \ref{fig:cavit}).  Physically, this is because the
toroidal magnetic field can escape via the polar jet before it
accumulates sufficiently in the magnetar wind bubble to drive the
termination shock to small radii (as occurs in 1D).  As a result, the
spindown of the neutron star is essentially identical to that of a
free wind, i.e., a wind without a surrounding stellar envelope
(Fig. \ref{fig:losses}).  For our lowest spindown power simulation,
corresponding to a 3 ms rotator, the termination shock does collapse
inside the fast magnetosonic surface (as \citealt{kom07} found in
their 2D simulations).  Even in this case, however, the neutron star's
spindown is still quite similar to that produced by a free wind, with
at most $\sim 30 \%$ differences (Fig. \ref{fig:losses}).  In
practice, these results imply that free wind calculations (e.g.,
\citealt{met07}) can be usefully used to study the winds from newly
formed magnetars.

\section*{Acknowledgments}
NB was supported by NASA through Hubble Fellowship grant
HST-HF-01193.01-A, awarded by the Space Telescope Science Institute,
which is operated by the Association of Universities for Research in
Astronomy, Inc., for NASA, under contract NAS 5-26555.  EQ was
supported in part by the David and Lucile Packard Foundation and
NSF-DOE Grant PHY-0812811. JA was supported by NSF grant AST-0507813,  
NASA grant NNG06G108G and DOE grant DE-FC02-06ER41453, all to the 
University of California, Berkeley, and by the taxpayers of California.

\vspace{-0.75cm}

%%%%%%%%%%%%%%%%%%%%%%%%%%%%%%%%%%%%%%%%%%%%%%%%%%%%%%%%%%%%%%%%%%

\label{lastpage}


\begin{thebibliography}{99}

\bibitem[\protect\citeauthoryear{Arons}{2003}]{aro03}
 Arons, J.\ 2003, \apj, 589, 871 

\bibitem[Begelman(1998)]{begelman98} 
Begelman, M.~C.\ 1998, \apj, 493, 291 

\bibitem[\protect\citeauthoryear{Begelman \& Li}{1992}]{beg92}
Begelman, M.~C., \& Li, Z.\ 1992, \apj, 397, 187

\bibitem[\protect\citeauthoryear{Berger et a.}{2003}]{ber03}
E. Berger, S. R. Kulkarni, D. A. Frail, \& A. M. Soderberg, 2003, ApJ, 599, 408 

\bibitem[\protect\citeauthoryear{Bucciantini et al.}{2006}]{me06}
 Bucciantini, N., Thompson, T.~A., Arons, J., Quataert, E., \& Del
 Zanna, L.\ 2006, \mnras, 368, 1717

\bibitem[\protect\citeauthoryear{Bucciantini et al.}{2007}]{b08a}
Bucciantini, N., Quataert, E., Arons, J., Metzger, B.~D., \&
Thompson, T.~A.\ 2007, \mnras, 380, 1541

\bibitem[\protect\citeauthoryear{Bucciantini et al.}{2008}]{b08b}
Bucciantini, N., Quataert, E., Arons, J., Metzger, B.~D., 
\& Thompson, T.~A.\ 2008, \mnras, 383, L25 

\bibitem[\protect\citeauthoryear{Burrows \& Lattimer}{1986}]{bur86}
 Burrows, A., \& Lattimer, J.~M.\ 1986, \apj, 307, 178 

\bibitem[\protect\citeauthoryear{Della Valle}{2006}]{del06}
 Della Valle, M.\ 2006, Chinese Journal of Astronomy and Astrophysics Supplement, 6, 010000 

\bibitem[\protect\citeauthoryear{Del Zanna et al.}{2007}]{ldz07}
 Del Zanna, L., Zanotti, O., Bucciantini, N., \& Londrillo, P.\ 2007, \aap, 473, 11 

\bibitem[\protect\citeauthoryear{Del Zanna \& Bucciantini}{2002}]{ldz02}
 Del Zanna, L., \& Bucciantini, N.\ 2002, \aap, 390, 1177 

\bibitem[\protect\citeauthoryear{Del Zanna et al.}{2003}]{ldz03}
 Del Zanna, L., Bucciantini, N., \& Londrilo, P.\ 2003, \aap, 400, 397 

\bibitem[\protect\citeauthoryear{Del Zanna et al.}{2004}]{ldz04}
 Del Zanna, L., Amato, E., \& Bucciantini, N.\ 2004, \aap, 421, 1063

\bibitem[Duncan \& Thompson(1992)]{duc92}
 Duncan, R.~C., \& Thompson, C.\ 1992, \apjl, 392, L9 

\bibitem[\protect\citeauthoryear{Frail et al.}{2001}]{fra01}
 Frail, D.~A., et al.\ 2001, \apjl, 562, L55 

\bibitem[\protect\citeauthoryear{Kennel \& Coroniti}{1984}]{ken84}
 Kennel, C.~F., \& Coroniti, F.~V.\ 1984, \apj, 283, 694  

\bibitem[Kohri et al.(2005)]{kohri05} 
Kohri, K., Narayan, R., \& Piran, T.\ 2005, \apj, 629, 341 

\bibitem[\protect\citeauthoryear{Komissarov \& Lyubarsky}{2004}]{kom04}
 Komissarov, S.~S., \& Lyubarsky, Y.~E.\ 2004, \mnras, 349, 779 

\bibitem[\protect\citeauthoryear{Komissarov \& Barkov}{2007}]{kom07}
Komissarov, S.~S., \& Barkov, M.~V.\ 2007, \mnras, 382, 1029 

\bibitem[K{\"o}nigl \& Granot(2002)]{kg02}
 K{\"o}nigl, A., \& Granot, J.\ 2002, \apj, 574, 134  

\bibitem[\protect\citeauthoryear{Lyubarsky \& Eichler}{2001}]{le01} 
Lyubarsky, Y.~E., \& Eichler, D.\ 2001, ApJ, 562, 494 

\bibitem[\protect\citeauthoryear{Lyutikov \& Blandford}{2003}]{lb03}
 Lyutikov, M., \& Blandford, R.\ 2003, ArXiv e-prints,
 arXiv:astro-ph/0312347

\bibitem[\protect\citeauthoryear{Maeda et al.}{2002}]{mae02}
 Maeda, K., Nakamura, T., Nomoto, K., Mazzali, P.~A., Patat, F., \&
 Hachisu, I.\ 2002, \apj, 565, 405 

%\bibitem[\protect\citeauthoryear{Maeda et al.}{2007}]{mae07}
% Maeda, K., et al.\ 2007, \apjl, 658, L5 

\bibitem[\protect\citeauthoryear{MacFadyen \& Woosley}{1999}]{mcf99}
 MacFadyen, A.~I., \& Woosley, S.~E.\ 1999, \apj, 524, 262 

\bibitem[\protect\citeauthoryear{McKinney \& Blandford}{2008}]{bla08}
McKinney, C.~J., \& Blandford, R.~D.\ 2008, astro-ph, arXiv:0812.1060

\bibitem[\protect\citeauthoryear{Matzner}{2003}]{mat03}
 Matzner, C.~D.\ 2003, \mnras, 345, 575 

\bibitem[\protect\citeauthoryear{Mazzali et al.}{2006}]{maz06}
 Mazzali, P.~A., et al.\ 2006, \nat, 442, 1018 

\bibitem[\protect\citeauthoryear{Nagataki et al.}{2003}]{nag03}
 Nagataki, S., Mizuta, A., Yamada, S., Takabe, H., \& Sato, K.\ 2003, \apj, 596, 401 

\bibitem[\protect\citeauthoryear{Nagataki et al.}{2006}]{nag06}
 Nagataki, S., Mizuta, A., \& Sato, K.\ 2006, \apj, 647, 1255

\bibitem[\protect\citeauthoryear{Metzger et al.}{2007}]{met07}
 Metzger, B.~D., Thompson, T.~A., \& Quataert, E.\ 2007, ApJ, 659, 561

\bibitem[\protect\citeauthoryear{Panaitescu \& Kumar}{2002}]{pan02}
 Panaitescu, A., \& Kumar, P.\ 2002, \apj, 571, 779 

\bibitem[Proga et al.(2003)]{proga03} Proga, D., MacFadyen, 
A.~I., Armitage, P.~J., \& Begelman, M.~C.\ 2003, \apjl, 599, L5 

\bibitem[\protect\citeauthoryear{Rhoads}{1999}]{rho99}
 Rhoads, J.~E.\ 1999, \apj, 525, 737

\bibitem[Soderberg et al.(2006)]{soderberg06} Soderberg, A.~M., 
Nakar, E., Berger, E., \& Kulkarni, S.~R.\ 2006, \apj, 638, 930 

\bibitem[Soderberg(2006)]{soderberg06b} 
Soderberg, A.~M.\ 2006, Gamma-Ray Bursts in the Swift Era, 836, 380 

\bibitem[\protect\citeauthoryear{Scheck et al.}{2006}]{sch06}
 Scheck, L., Kifonidis, K., Janka, H.-T., M\"{u}ller, E.\ 2006, \aap, 457, 963 

\bibitem[\protect\citeauthoryear{Sherwin \& Lynden-Bell}{2007}]{she07}
 Sherwin, B.~D., \& Lynden-Bell, D.\ 2007, \mnras, 378, 409

\bibitem[\protect\citeauthoryear{Thompson}{1994}]{thom94} 
Thompson, C.\ 1994, MNRAS, 270, 480 

\bibitem[Thompson et al.(2001)]{thompson01}
 Thompson, T.~A., Burrows, A., \& Meyer, B.~S.\ 2001, \apj, 562, 887 

\bibitem[\protect\citeauthoryear{Thompson et al.}{2004}]{thom04}
Thompson, T.~A., Chang, P., Quataert, E.\ 2004, ApJ, 611, 380

\bibitem[\protect\citeauthoryear{Thompson et al.}{2005}]{thom05}
Thompson, T.~A., Quataert, E., \& Burrows, A.\ 2005, \apj, 620, 861

\bibitem[\protect\citeauthoryear{Tomimatsu et al.}{2001}]{tomi01}
 Tomimatsu, A., Matsuoka, T., \& Takahashi, M.\ 2001, Phys. Rev. D, 64, 123003 

\bibitem[\protect\citeauthoryear{Usov}{1992}]{usov92} 
Usov, V.~V.\ 1992, Nature, 357, 472 

\bibitem[\protect\citeauthoryear{Uzdensky \& MacFadyen}{2006}]{um06}
 Uzdensky, D.~A., \& MacFadyen, A.~I.\ 2006, \apj, 647, 1192

\bibitem[\protect\citeauthoryear{Uzdensky \& MacFadyen}{2007}]{um07} 
Uzdensky, D. A. \& MacFadyen, A. I., 2007, ApJ accepted (astro-ph/0609047)

\bibitem[\protect\citeauthoryear{Wheeler et al.}{2000}]{wheeler00} 
Wheeler, J.~C., Yi, I., H{\"o}flich, P., \& Wang, L.\ 2000, ApJ, 537, 810 

\bibitem[\protect\citeauthoryear{Wheeler et al.}{2008}]{whe08}
 Wheeler, J.~C., Maund, J.~R., \& Couch, S.~M.\ 2008, \apj, 677, 1091 

\bibitem[\protect\citeauthoryear{Woosley \& Bloom}{2006}]{woo06}
 Woosley, S.~E., \& Bloom, J.~S.\ 2006, ARA\&A, 44, 507

\bibitem[\protect\citeauthoryear{Woosley et al.}{2002}]{woo02}
 Woosley, S.~E., Heger, A., \& Weaver, T.~A.\ 2002, Reviews of Modern
 Physics, 74, 1015

\bibitem[\protect\citeauthoryear{Woosley \& Weaver}{1995}]{ww95}
Woosley, S.~E., \& Weaver, T.~A.\ 1995, ApJS, 101, 181

\bibitem[\protect\citeauthoryear{Zhang}{2007}]{zha07}
 Zhang, B.\ 2007, Chinese Journal of Astronomy and Astrophysics, 7, 1 

\end{thebibliography}
\end{document}